# Superfluid Spin Transport in the Van der Waals Antiferromagnet $CrCl_3$

Peisen Yuan[1,2,5*], Xiaomin Guo[1,5], Vincent Flynn[3], Benedetta Flebus[3], Fèlix Casanova[1,4*], and Luis E. Hueso[1,4*]

[1] *CIC nanoGUNE BRTA, 20018 Donostia-San Sebastian, Basque Country, Spain*

[2] *State Key Laboratory of Flexible Electronics (LoFE) & Institute of Flexible Electronics (IFE), Xiamen University, Xiamen 361102, China*

[3] *Department of Physics, Boston College, Chestnut Hill, Massachusetts 02467, United States*

[4] *IKERBASQUE, Basque Foundation for Science, 48009 Bilbao, Basque Country, Spain*

[5] *These authors contributed equally to this work.*

*Email: y.peisen@nanogune.eu, f.casanova@nanogune.eu, l.hueso@nanogune.eu

**Over the past decade, the quest for spin superfluidity has moved to the forefront of spintronics, driven by the promise of phase-gradient-driven, ultra-low-loss spin transport. However, experimental investigations remain limited, primarily due to the lack of suitable material systems. Here, we report on the discovery and control of a superfluid spin transport in the easy-plane van der Waals antiferromagnetic (AFM) insulator $CrCl_3$ by a nonlocal device structure. Combining nonlocal magnon transport measurements with theoretical modelling, we demonstrate that spin superfluidity emerges in $CrCl_3$ under canted AFM spin configurations, where it gives rise to ultra-long range (around 90 μm), weakly decaying spin transport. We also provide direct evidence that strong magnetic fields and elevated temperatures suppress the superfluid state, restoring the rapid exponential decay with distance of incoherent magnons. These findings underscore the potential of spin superfluidity in two-dimensional magnetic insulators and establish $CrCl_3$ as a promising platform for energy-efficient, long-distance spin transport in next-generation spintronic applications.**

Supercurrents -phase-coherent flows immune to ordinary scattering- have been a central pursuit across physics, as their coherence can convert microscopic order into macroscopic functionalities. In superconductors, supercurrents of charge enable lossless transport and long-range quantum coherence on macroscopic scales, underpinning technologies from power transmission to quantum processors[1]. In superfluids, persistent flows of mass exemplify collective quantum order and inspire analogues of gravitational and cosmological phenomena[2-8]. Extending this paradigm to other platforms has led to the identification of spin superfluidity, i.e., an emergent, phase-coherent regime of spin-angular-momentum transport in magnetic insulators (MIs) [9]. In principle, spin superfluids can generate spin transport without the exponential attenuation characteristic of diffusive magnons, holding thus great promise for enabling next-generation, ultra-low-power spintronic technologies. A nonlocal metal/MI/metal device geometry has been established as electrical platform for generating and detecting spin currents by employing metallic contacts as spin injectors and detectors[10-15]. Theoretically, such architectures have been predicted to be promising probes of spin superfluidity[16-19]. Yet, despite years of theoretical proposals, clear experimental realization and control of spin superfluidity has been elusive, owing to the ubiquitous presence of damping, anisotropies, and parasitic interactions that destabilize its coherence.

In this work, we experimentally demonstrate the emergence of spin superfluidity in a van der Waals (vdW) antiferromagnetic insulator (AFI), chromium trichloride ($CrCl_3$), enabling long-distance spin supercurrent transport characterized by a remarkably slow decay of magnon signals in a nonlocal device geometry. In agreement with our theoretical model, the spin superfluid state arises in $CrCl_3$ when it is in a canted antiferromagnetic (AFM) magnetic configuration. A strong enough magnetic field or high enough temperature suppresses this state, either by inducing a fully aligned spin configuration or by destroying the conditions necessary for spin superfluidity, leading to the conventional exponential decay of magnon currents.

VdW magnets have recently emerged as prime candidates for realizing spin superfluidity, owing to their tunable layer-dependent magnetic properties, including anisotropy, interlayer exchange, and the emergence of topological spin textures[20-26]. Among this class of materials, $CrCl_3$ is particularly promising[25]. $CrCl_3$ is an electrical insulator that undergoes a paramagnetic to AFM phase transition upon cooling with a Néel temperature ($T_N$) of approximately 14 K[27]. The exceptionally weak interlayer AFM exchange interaction (~1.6 μeV) and vanishingly small in-plane magnetic anisotropy (10 Oe) in $CrCl_3$ facilitates the manipulation of its spin state under relatively low external magnetic fields[28-30]. In monolayers, theoretical and experimental studies point to meron-antimeron textures[25,31,32], highlighting the near-U(1) easy-plane character and low pinning of its magnetization. These same ingredients can in principle support phase-coherent transport of the Néel order in multilayer crystals. In addition, $CrCl_3$ exhibits a relatively low spin-flop transition ($B_{sf}$~160 Oe) and saturation ($B_{sat}$~0.17 T) fields in bulk crystals[28], making it a promising platform for exploring the pure spin current transport across different magnetic phases. Notably, in exfoliated $CrCl_3$ flakes, the interlayer exchange coupling is enhanced with decreasing thickness, leading to a 5-10-fold increase in $B_{sf}$ and $B_{sat}$ compared to those of bulk crystals[24,33].

Figure 1a shows the crystal structure of $CrCl_3$, in which each $Cr^{3+}$ ion is octahedrally coordinated by six $Cl^-$ ions. The Raman spectra of bulk and exfoliated $CrCl_3$ flakes are present in Figure 1b. Six characteristic Raman resonant modes are observed in the bulk sample; however, several modes diminish or vanish as the thickness decreases, consistent with recent reports on thickness-dependent phonon behavior in $CrCl_3$[34,35]. The weak vdW interlayer bonding allows the mechanical exfoliation of large-area, atomically thin flakes[27,29], facilitating the fabrication and investigation of long-distance spin transport in nonlocal device geometries.

To investigate the spin current transport properties of $CrCl_3$, we employ a nonlocal Pt/$CrCl_3$/Pt device structure consisting of multiple top-contact Pt strips with varying center-to-center distance ($d$) (Figure 1c and 1d). Nonlocal spin transport using this type of device structure has been previously demonstrated in several vdW AFIs[36-39]. In this configuration, magnons can be excited in the $CrCl_3$ layer either electrically or thermally via current injection through one of the Pt strips[10-12,40,41]. For electrical excitation, a current $I$ passing through a Pt injector strip generates a transverse spin current at the Pt/MIs interface via spin Hall effect (SHE) [10,42,43]. This process induces a local spin accumulation that injects magnons into $CrCl_3$, bringing the electrically-excited magnon current through $CrCl_3$[44]. Meanwhile, the Pt injector also serves as a local heater due to Joule heating, creating a radial temperature gradient ($\nabla T$) that drives a thermally-excited magnon current in the $CrCl_3$ layer via the spin Seebeck effect[45]. Beyond the standard incoherent magnon currents, at sufficiently strong injection, magnon-magnon interactions transfer angular momentum from the incoherent thermal-magnon cloud into a phase-coherent superfluid[46,47], which then provides the primary current path in $CrCl_3$ (Figure 1c). If the conditions for stabilizing the superfluid are not met, spin transport is instead carried by diffusive thermal magnons. The propagating spin current (carried by either normal magnons or coherent spin supercurrent) is subsequently detected at a spatially separated Pt strip through the inverse spin Hall effect (ISHE), leading to the nonlocal first-harmonic ($V_{NL}^{1\omega}$, electrical excitation) and second-harmonic ($V_{NL}^{2\omega}$, thermal excitation) voltage signal, respectively[10,41,48-50]. In this study, our analysis primarily focuses on the $V_{NL}^{2\omega}$ signal since it exhibits robust and reproducible behavior across all devices. In contrast, the $V_{NL}^{1\omega}$ signal is extremely weak and detected only in very thick $CrCl_3$ flakes (over 80 nm) with very short $d$ (sub μm), making unsuitable for a systematic investigation of spin-transport behavior in our device geometry.

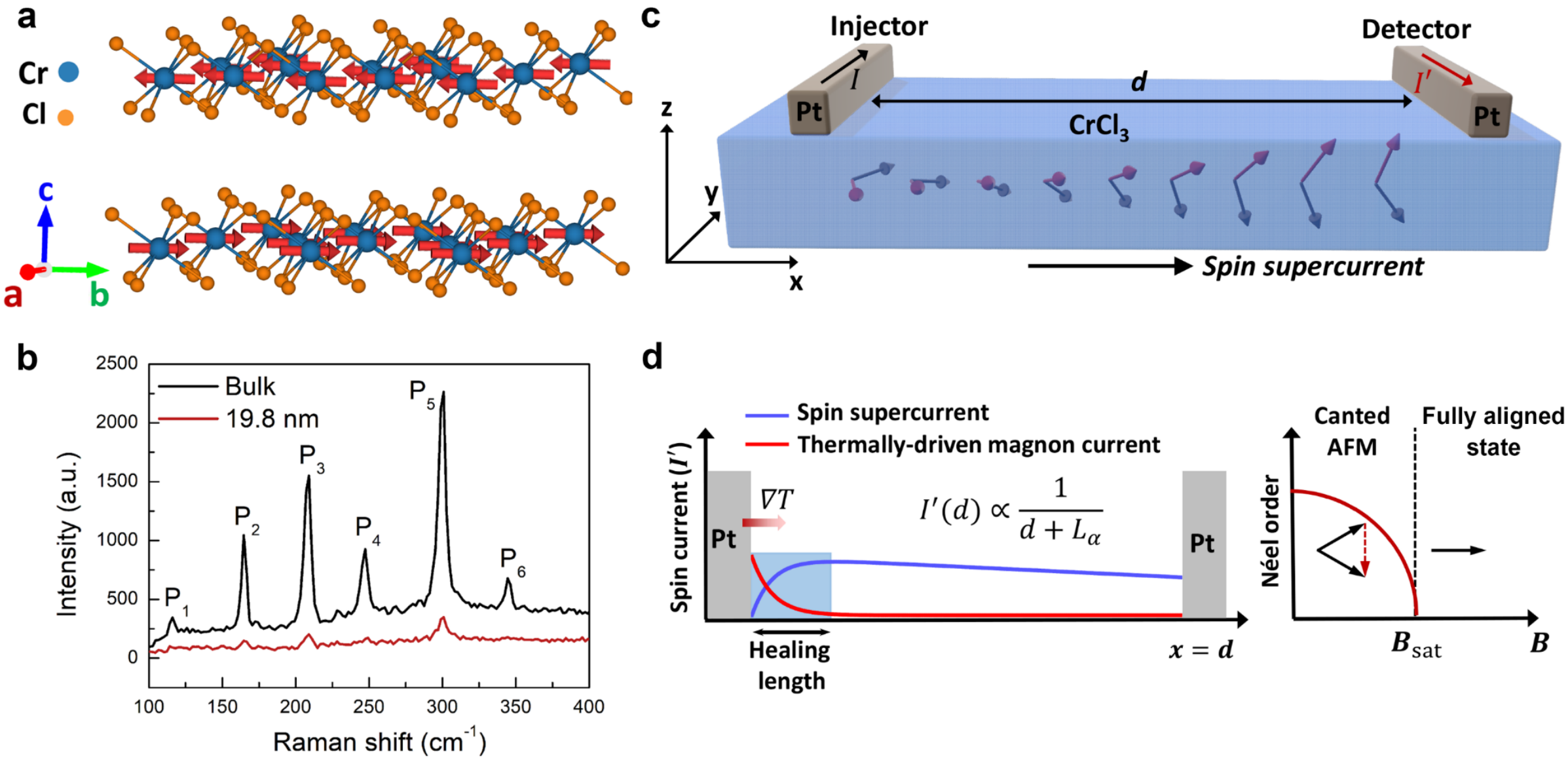

**Figure 1. Superfluid spin transport in $CrCl_3$ within a lateral device.** **a**, Crystal and magnetic structure of $CrCl_3$. **b**, Raman spectra comparing bulk $CrCl_3$ and an exfoliated flake used in the fabricated lateral device (Device 1). **c**, Schematic illustration of a coherent phase gradient in the spin precession angle, which drives the flow of a spin supercurrent in a nonlocal transport geometry. **d**, Left: Schematic of a nonlocal spin current transport in the Pt/$CrCl_3$/Pt system. A charge current applied in the left Pt contact creates a temperature gradient at the interface, causing an injection of spin current into the adjacent $CrCl_3$ via the spin Seebeck effect. Thermally-driven magnons scatter and dissipate, ultimately condensing into the spin superfluid state and efficiently propagating across the device. This transfer of spin angular momentum occurs over a characteristic healing length (see Section 1 of the Supporting Information), assumed to be much smaller than the device length $d$. Due to intrinsic damping, the spin-supercurrent decays algebraically with $d$. At the right Pt contact, the transmitted spin current is converted into a measurable voltage via the ISHE. Right: Evolution of the Néel order parameter with applied magnetic field ($B$). For $B < B_{sat}$, adjacent $CrCl_3$ layers form a canted AFM ground state with finite Néel order (quantified by the Néel vector magnitude, shown as dashed red arrow). For $B > B_{sat}$, the Néel order vanishes as magnetic moments becomes fully polarized, quenching spin superfluidity.

In order to confirm the possible realization of superfluid spin transport in $CrCl_3$ within our device geometry, we develop a theoretical framework that describes the generation and transport of a superfluid spin current in a two-sublattice antiferromagnet with easy-plane anisotropy. The model (see Section 1 of the Supporting Information for a detailed description) reveals that the canted AFM phase supports a nearly U(1)-symmetric free energy landscape, permitting low-energy, long-wavelength phase fluctuations of the Néel vector. Although a weak easy-plane anisotropy breaks this symmetry and suppresses phase coherence in equilibrium, the spin superfluid state can nevertheless be dynamically sustained when the injection of sufficiently strong spin angular momentum -carried by thermal magnons at the interface- overcomes the threshold set by these parasitic terms. Within the range of fields supporting nonlocal transport, we do not observe a sharp loss in signal below a given applied current threshold. This suggests that the current threshold lies below the resolution of our measurement devices. In this regime, the presence of a spin superfluid state in $CrCl_3$ leads to a markedly slower decay of the spin current ($I'$) as a function of $d$, as illustrated in Figure 1d and expressed by the following relation:

$$I' \propto \frac{1}{d+L_\alpha} \quad (1)$$

where $L_\alpha$ is a phenomenological constant related to the Gilbert damping of the material and the interfacial spin-mixing conductance of the Pt/$CrCl_3$ interface (see Eq. (8) in the Supporting Information for the precise relationship). The resulting long-range and weakly decaying of superfluid spin transport in $CrCl_3$ can be experimentally verified using a nonlocal device geometry, as detailed in the following. Crucially, our model predicts that spin superfluidity is sustained only at low magnetic fields and temperatures, where the canted AFM state enables a phase gradient and thermal fluctuations do not destroy phase coherence. Independently of temperature, at higher fields the system saturates to a fully aligned state, eliminating the Néel degree of freedom and suppressing spin supercurrents (Figure 1d).

Experimentally, we first investigate the spin transport signal using a lateral device structure with a fixed $d$. Figure 2a exhibits an optical image of the Pt/$CrCl_3$/Pt device along with the measurement configuration used to detect both local and nonlocal spin signals (see Methods for details). A series of lateral Pt/$CrCl_3$/Pt devices (Devices 1-6) with varying thicknesses were prepared for this study (Supplementary Figure 1). Efficient generation and detection of local spin signals in Pt/$CrCl_3$ heterostructures have been confirmed via spin Hall magnetoresistance (SMR) (see Section 2 of the Supporting Information, and Supplementary Figure 2 and 3). Figure 2b displays the angle-dependent magnetoresistance (ADMR) measurements of $V_{NL}^{2\omega}$ at a fixed magnetic field ($B$=0.7 T) under different injection current ($I$) in Device 1. All curves are well described by a $\cos\theta$ function, where the angle between $B$ and $I$ is defined as $\theta$, with $\theta$ = 0° when $B$ and $I$ are perpendicular. The inset of Figure 2b plots the current dependence of the amplitude of $V_{NL}^{2\omega}$ ($\Delta V_{NL}^{2\omega}$), extracted from these ADMR measurements. A clear linear relationship between $\Delta V_{NL}^{2\omega}$ and $I^2$ is observed for both low (0.7 T) and high fields (4 T), consistent with the thermally driven nature of magnon excitation.

Figure 2c presents the ADMR measurements of $V_{NL}^{2\omega}$ at fixed $I$ (10 μA) under different applied $B$. The temperature dependence of the $\Delta V_{NL}^{2\omega}$ (inset of Figure 2c) shows that the signal persists up to ~13 K, close to the $T_N$ of $CrCl_3$, confirming its magnetic origin. Control experiments exclude artifacts such as charge carrier transport (Supplementary Figure 4). No nonlocal voltage is observed in $Pt/SiO_x/Pt$ devices under similar measurement conditions (Supplementary Figure 5), confirming that the detected signals originate from spin transport in $CrCl_3$. Furthermore, the $V_{NL}^{2\omega}$ signal exhibits only a slight reduction after the sample is stored under ambient conditions for one week (Supplementary Figure 6), indicating that the spin transport in $CrCl_3$ is relatively insensitive to degradation from oxygen and moisture. Notably, although all ADMR curves exhibit a $cos\theta$ dependence at both low and high fields, their underlying physical origins are distinct. At magnetic fields (e.g., 2-4 T) above $B_{sat}$ (around 1.9 T for our devices), all magnetic moments in $CrCl_3$ align with the field direction, yielding a net moment $m$ along $B$. In this case, the spin superfluid state is suppressed, and the detected $V_{NL}^{2\omega}$ arises solely from the conventional thermally-induced magnon transport. For diffusive magnon transport in AFIs, it is reported that the nonlocal signal is influenced by both the Néel vector $n$ and the field-induced net magnetic moment $m$[11]. Specifically, thermally-driven magnons are predominantly proportional to the component of $m$ that is perpendicular to the Pt strip. As $B$ is rotated, the projection of $m$ normal to the Pt strip follows a $cos\theta$ profile, resulting in the corresponding angular dependence in $V_{NL}^{2\omega}$, with extrema at $\theta = 0°$ and 180°.

At low magnetic fields, where spin superfluidity is expected to emerge, the measured $V_{NL}^{2\omega}$ still exhibits a clear $cos\theta$ angular dependence (Figure 2c). This behavior is consistent with theoretical predictions that the spin supercurrent arriving at the $Pt/CrCl_3$ interface can be pumped into Pt and detected by the ISHE[16-18], thereby preserving the characteristic $cos\theta$ dependence. After normalizing $\Delta V_{NL}^{2\omega}$ by the length of the Pt strip, we obtain an average nonlocal magnon signal exceeding $10^8$ $V/(A^2 \cdot m)$. This value is 1 to 3 orders of magnitude higher than those reported for other vdW or even three-dimensional (3D) MIs, as summarized in Supplementary Table S1[10,11,36-39,51-53]. The exceptionally large magnon currents observed in such thin $CrCl_3$ layers create favorable conditions for the formation of a spin superfluid state at low temperature, thereby facilitating the onset of superfluid spin transport.

Figure 2d presents the field-dependent magnetoresistance (FDMR) measurement of $V_{NL}^{2\omega}$ at $\theta$ =0° and 90°. When $B$ is oriented perpendicular to $I$ ($\theta$ =0°), $V_{NL}^{2\omega}$ increases with increasing $B$ and saturates beyond 1.9 T. This behavior is consistent with the local FDMR measurement results (Supplementary Figure 3g and 3h), indicating the same $B_{sat}$ (1.9 T) of $CrCl_3$. Regarding the spin-flop transition, previous studies have reported a thickness-dependent reduction, with $B_{sf}$ value decreasing as the $CrCl_3$ layer becomes thicker[33]. In our devices, all with thicknesses greater than 8 nm, $B_{sf}$ is expected to fall below 0.2 T, which is consistent with our observation of finite $V_{NL}^{2\omega}$ signals at fields lower than 0.2 T (Figures 2c and 2d). In contrast, when $B//I$ (θ =90°), the $V_{NL}^{2\omega}$ signal is zero across the entire field range. This is expected, as the component of the $m$ normal to the Pt strip vanishes in this configuration. Notably, high injection currents significantly impact the nonlocal magnon signal (Supplementary Figure 7), a behavior also observed in other vdW AFIs[36,53]. The linear $\Delta V_{NL}^{2\omega}$-$I^2$ relation holds only within a current range, with deviations above 10 μA (Supplementary Figure 7a) which we attribute to the increase of the base temperature and the suppression of magnetic order. This is further supported by reduced magnetic features in high-current FDMR curves (Supplementary Figure 7b). To minimize overheating effects, all subsequent measurements are performed with $I \leq 10$ μA. We also explored the $V_{NL}^{1\omega}$ signals (electrical excitation) in our $CrCl_3$ devices (Section 3 in the Supporting Information and Supplementary Figure 8-10). However, electrically injected magnon currents are significantly less efficient than thermally generated magnons. Moreover, with increasing temperature, decreasing sample thickness, and increasing channel length ($d$), the already weak $V_{NL}^{1\omega}$ signal is further suppressed and readily falls below the instrumental detection limit. In addition, as discussed in the theoretical section (Section 1 in the Supplementary Information), the finite easy-plane magnetic anisotropy of $CrCl_3$ requires a sufficiently large magnon population to overcome the anisotropy-induced pinning and activate coherent spin dynamics. Consequently, weak electrical magnon excitation may be insufficient to effectively convert incoherent magnons into the spin superfluidity capable of long-range transport. Under such conditions, the confidence in extracting meaningful spin-transport trends from the first-harmonic response is substantially reduced. For these reasons, we focused our study on thermally generated magnon currents to investigate spin current transport in $CrCl_3$.

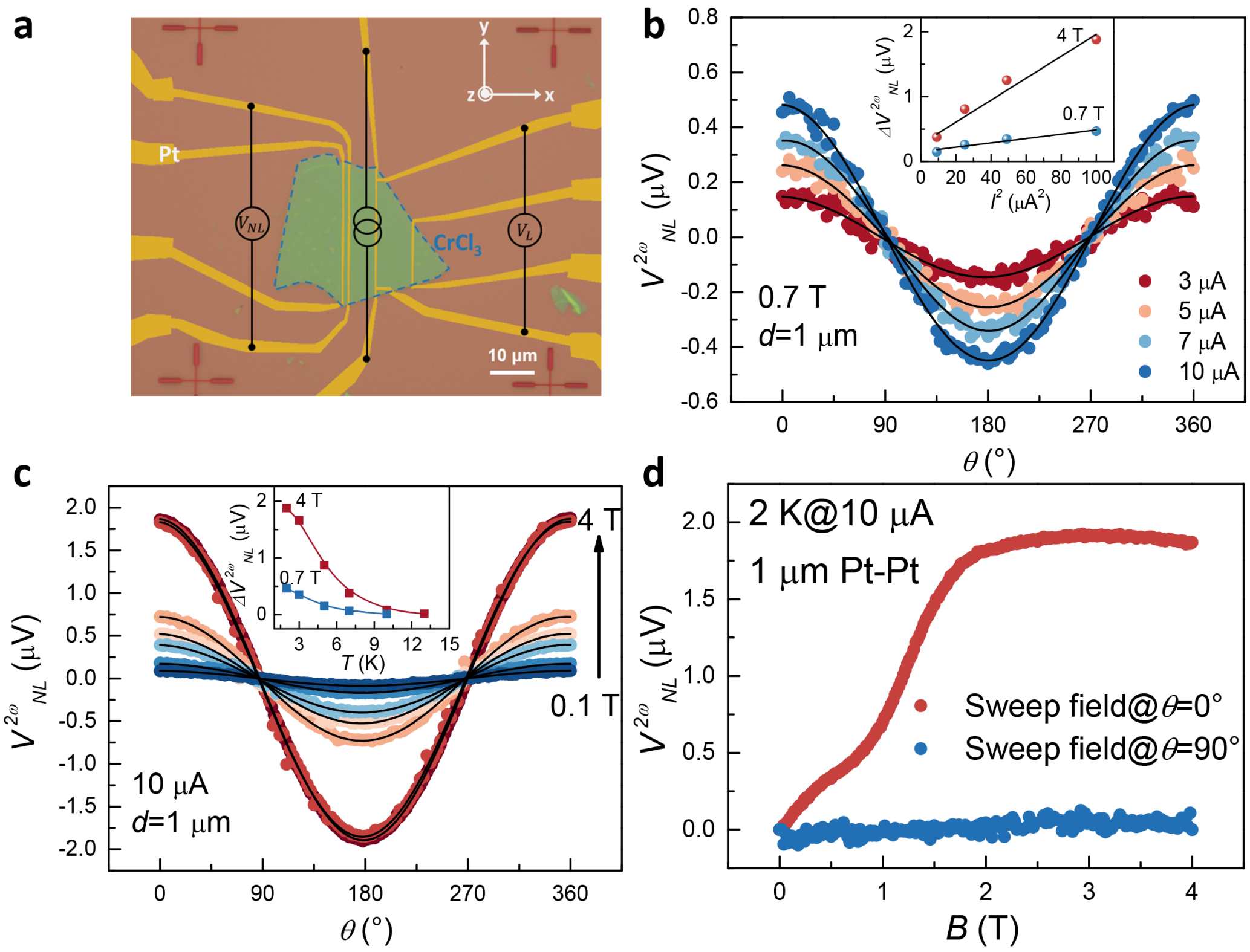

**Figure 2. Nonlocal magnon transport measurement in Device 1. a**, Optical microscope image of the fabricated nonlocal device with a 19.8-nm-thick $CrCl_3$. An in-plane (IP) magnetic field $B$ is applied at an angle $\theta$ relative to the injected current $I$, where $\theta = 0°$ corresponds to $B$ perpendicular to $I$. **b**, ADMR measurements of $V_{NL}^{2\omega}$ measured at $B$=0.7 T under different $I$. Black solid lines indicate fits to a $cos\theta$ dependence. Inset: extracted amplitude $\Delta V_{NL}^{2\omega}$ plotted as a function of the square of the injection current ($I^2$) at two different magnetic fields (0.7 T and 4 T). **c**, ADMR of $V_{NL}^{2\omega}$ measured under different magnetic fields. Black solid lines represent $cos\theta$ fits. Inset: temperature dependence of the ADMR amplitude, $\Delta V_{NL}^{2\omega}$, at selected magnetic fields. **d**, Field-dependent measurements (FDMR) of $V_{NL}^{2\omega}$ for two configurations: $B$ perpendicular ($\theta = 0°$) and parallel ($\theta = 90°$) to $I$.

The occurrence of superfluid spin transport inside $CrCl_3$ can be directly checked by analyzing $\Delta V_{NL}^{2\omega}$ as a function of $d$ in lateral devices. As shown in Figures 3a and 3b, the plots of $\Delta V_{NL}^{2\omega}$ versus $d$ for Device 2 (8 nm) reveal a significantly slower decay at low $B$ (below 1 T) compared to the typical exponential decay expected for diffusive magnon transport at high $B$ (above $B_{sat}$), which is described by[10,36]:

$$\Delta V_{NL}^{2\omega} = \frac{C}{\lambda_m}\frac{\exp\left(\frac{d}{\lambda_m}\right)}{1-\exp\left(\frac{2d}{\lambda_m}\right)} \tag{2}$$

where $\lambda_m$ is the magnon diffusion length and $C$ is a distance-independent constant. The slowly decaying behavior is reproducible across devices with varying $CrCl_3$ thicknesses, excluding possible artifacts (Supplementary Figure 11). In earlier studies on thermally-driven magnon transport in YIG, a slower magnon decay rate than the general exponential decaying has been attributed to the intrinsic spin Seebeck effect[48,49]. In that case, when $d \geq$3-5 $\lambda_m$, the detected nonlocal thermal voltage $V_{NL}^{2\omega}$ is no longer governed by $\mu_m$ induced by $\nabla T$ from the Pt heater but it is influenced by a small residual $\nabla T$ at the MI/substrate interface. This mechanism typically results in a decay behavior following a $1/d^2$ dependence. However, in our $CrCl_3$ devices, the experimental data do not satisfy a $1/d^2$ dependence (Figure 3b), thereby ruling out this intrinsic spin Seebeck effect as the origin of the slow magnon decay observed here. To further rule out the possible contribution of thermal gradients generated at the Pt injector to the observed nonlocal signals at the Pt detector, we systematically characterized the temperature profile along an 8-nm-thick $CrCl_3$ channel (Section 4 in the Supporting Information and Supplementary Figure 12-15). When a current of 10 μA is applied to the Pt injector, the temperature change at the detector position becomes negligible for d⩾2 μm. The nonlocal signals measured in this distance range (Figure 3a) therefore arise in the absence of any temperature variation at the detector, excluding any contribution of thermally induced effects to the long-range spin transport signal.

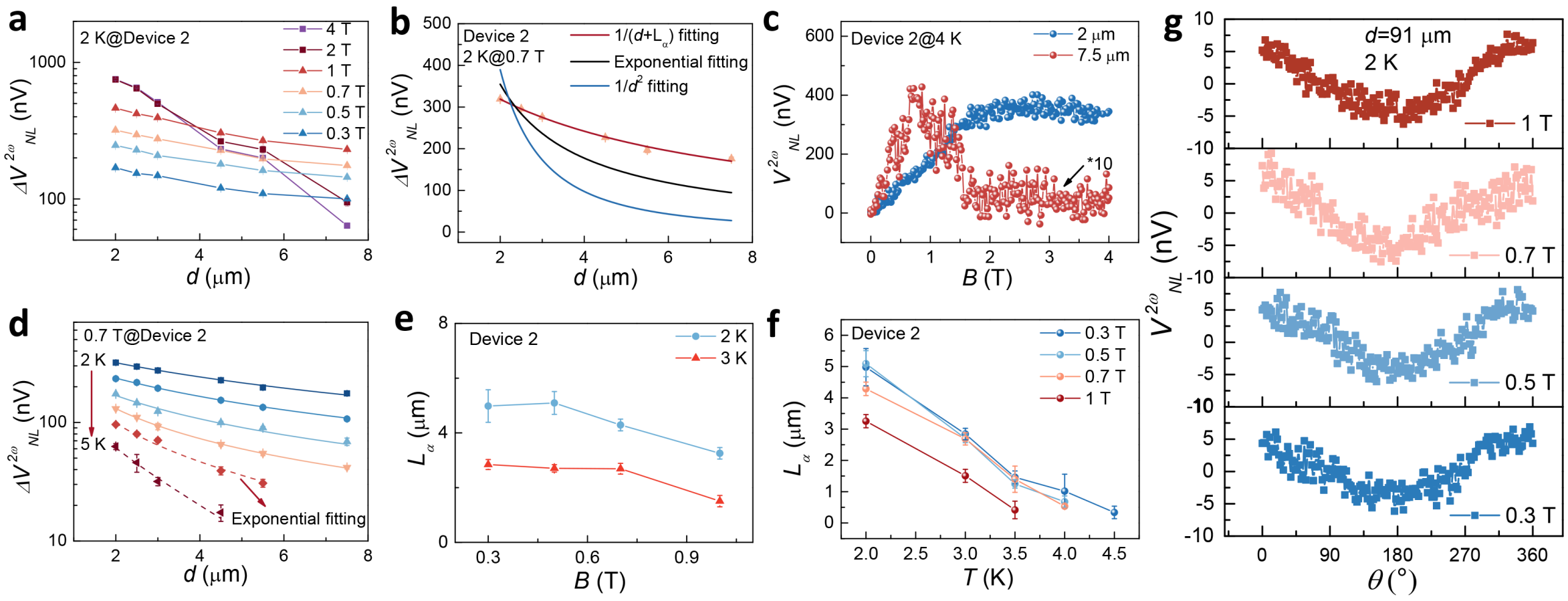

**Figure 3. Experimental evidence of superfluid spin transport in $CrCl_3$. a**, Semilogarithmic plot of $\Delta V_{NL}^{2\omega}$ as a function of $d$ at various applied magnetic fields at 2 K for Device 2. **b**, Comparison of the fitting of $\Delta V_{NL}^{2\omega}$ versus $d$ at 0.7 T and 2 K for Device 2 using different models. **c**, FDMR of $V_{NL}^{2\omega}$ of Device 2 at $d<\lambda_m$ and $d>\lambda_m$. In the case of 4 K, $\lambda_m$ of Device 2 is around 2 μm (see the following studies). **d**, $\Delta V_{NL}^{2\omega}$ versus $d$ at 0.7 T and different temperatures for Device 2. The solid and dashed lines represent the fitted curves by the superfluid spin transport model of Eq. 1 (2 K, 3 K, 3.5 K and 4 K) and a conventional exponential decay function (4.5 K and 5 K), respectively. **e**, Fitted $L_\alpha$ values as function of applied magnetic field at 2 K and 3 K for Device 2. f, The temperature dependence of $L_\alpha$ values at different $B$ for Device 2. **g**, Nonlocal spin signal detected over an ultra-long distance ($d$= 91 μm) at various applied magnetic fields in a 23-nm-thick $CrCl_3$ device.

Importantly, beyond a characteristic distance $d$, the low-field nonlocal voltage clearly exceeds that measured at high fields, exhibiting a pronounced crossover point (Figure 3a). Both the field-dependent transition in the decay behavior observed over the same range of $d$ and the emergence of this crossover cannot be accounted for conventional diffusive magnon transport. More strikingly, at sufficient long distances, we observed a sharp, field-driven onset and disappearance of the nonlocal signal (Figure 3c and Supplementary Figure 14f). Considering that the basic spin Hall conversion of charge current in Pt into a spin current (interfacial spin-mixing conductance, heating efficiency, etc.) is not expected to change by orders of magnitude over the field range of our experiment, so the injected spin current should be largely insensitive to $B$. By contrast, the underlying magnetic configuration changes qualitatively with field: only in the weak-field regime does the system support a stable precessing order parameter and hence a spin-superfluid solution. The observation that the long-range signal is quenched only by a large enough magnetic field is therefore most naturally interpreted as the system crossing in and out of the superfluid regime. We have also considered magnon Hanle transport as an alternative explanation for the peak-like field dependence.[54] However, Hanle pseudospin precession remains a diffusive process and therefore yields an exponentially damped, potentially oscillatory spatial dependence. It does and cannot naturally account for the observed algebraic transport regime and its field- and temperature-driven crossover to exponential decay.

Indeed, all slowly decaying curves are well described by Eq. 1 (Figures 3b and 3d, Supplementary Figure 16). Notably, the parameter $L_\alpha$ remains nearly constant over the magnetic field range of 0.3 to 0.7 T but decreases with increasing temperature (Figure 3e). These results are highly reproducible (see Supplementary Figure 17 for results in another sample). According to theoretical calculations, $L_\alpha$ is mainly related with the Gilbert damping of the material and the interfacial spin-mixing conductance (see the details in Section 1 in the Supporting Information and refs.16-19). Additionally, it is only weakly dependent on the applied magnetic field, with a dependence on the applied field on the leading order of $(B/B_{sat})^2$,where $B_{sat}{\sim}1.9\ T$ is the spin saturation field. In the weak-field regime $\ll B_{sat}$ , the characteristic length scale $L_\alpha$ can be shown to be largely insensitive to $B$ (see Section 1 in the Supporting Information). At higher temperatures, increased magnon-magnon scattering and enhanced Gilbert damping suppress the long-range phase coherence required for superfluid spin transport, leading to the reduction of $L_\alpha$ (Figure 3f). As a result, incoherent thermal magnons become the dominant spin carriers, and the algebraic decay of the supercurrent transitions to an exponential attenuation, as evidenced by our experimental results for a characteristic T=4.5 K (Figure 3d and Supplementary Figure 16). Furthermore, in a canted antiferromagnet, increasing the magnetic field $B$ progressively weakens the Néel order as the system approaches $B_{sat}$, thereby reducing the phase stiffness of the coherent spin texture. As a result, the characteristic temperature for sustaining superfluid spin transport decreases with increasing $B$, which agrees well with our experimental observations (Figure 3f). Additional evidence supporting superfluid spin transport in $CrCl_3$ is the persistence of $V_{NL}^{2\omega}$ signals at ultra long $d$ (Figure 3g and Supplementary Figure 18). In particular, the $V_{NL}^{2\omega}$ is only detected at low field regime (<1 T) at 91 μm (Supplementary Figure 18b). This result indicates the contribution of spin superfluidity for long-range spin signal transport, which also represents the longest spin current transport distance reported to date in magnetic systems using a nonlocal device geometry by thermally excited magnons.

The above results demonstrate that superfluid spin transport in $CrCl_3$ is stable only under low magnetic fields within the canted AFM phase. Once the external magnetic field becomes sufficiently strong to destroy the coherent long-range magnetic order, the spin superfluid state is suppressed, and spin transport crosses over to conventional thermal magnon diffusion, as evidenced by the clear transition in the decay behavior shown in Figure 3a. We next explore this conventional thermal magnon transport regime in $CrCl_3$. Figure 4a presents $\Delta V_{NL}^{2\omega}$ as a function of $d$ in Device 2 measured at 4 T, which is above $B_{sat}$. The data at various $CrCl_3$ thicknesses and temperatures are well described by Eq. 2 (Figure 4a and Supplementary Figure 19), allowing reliable extraction of $\lambda_m$. We further analyze the temperature dependence of $\lambda_m$ (inset of Figure 4a) and its thickness dependence (Figure 4b) in Pt/$CrCl_3$/Pt nonlocal devices. Notably, $\lambda_m$ increases as the temperature decreases. This behavior is consistent with previous studies on the relaxation of conventional thermal magnons in other AFIs[36,38,39]. One possible reason is the enhancement magnon lifetime at lower temperatures[36]. The thickness dependence of $\lambda_m$ shows a monotonic increase followed by saturation at thicker flakes, reaching a maximum value of 14 ± 3 μm at 2 K. This trend suggests that surface scattering is reduced by increasing thickness, so that $\lambda_m$ saturates once the flake is thick enough and surfaces no longer limit the magnon transport, approaching the intrinsic bulk value. Figure 4c summarizes the state-of-the-art maximum $\lambda_m$ values reported for several representative MIs, including both vdW and 3D materials.

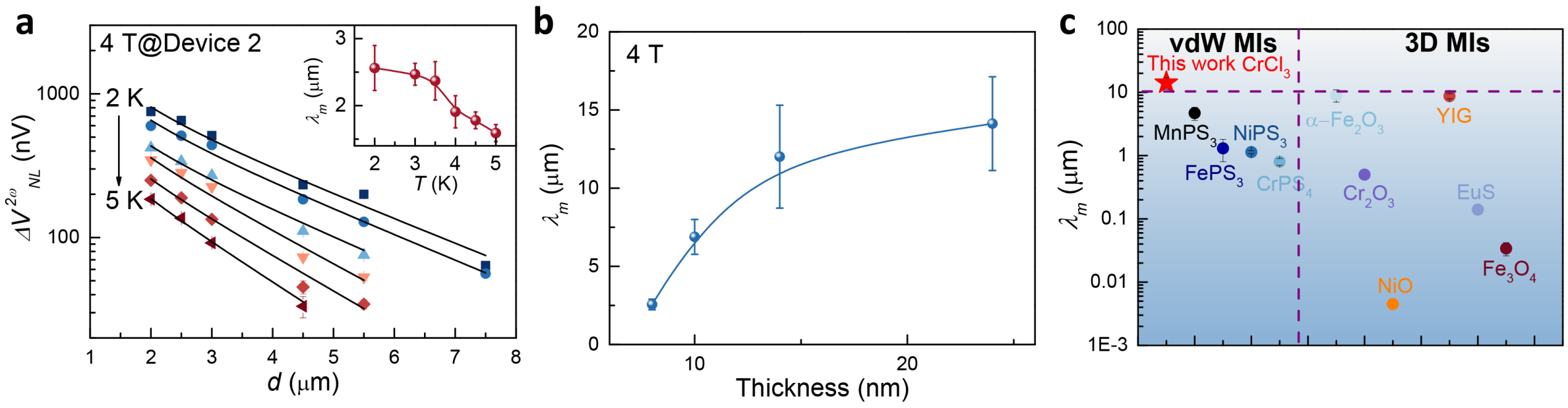

**Figure 4. Conventional diffusive magnon transport at the field-induced FM state of $CrCl_3$. a**, $\Delta V_{NL}^{2\omega}$ as a function of $d$ at different temperatures. Solid lines represent exponential fits based on Eq. 2 in the main text. Inset: temperature dependence of the extracted $\lambda_m$. **b**, Thickness dependence of $\lambda_m$ at 2 K. **c**, Comparison of state-of-art $\lambda_m$ values for various MIs.

For comparison, all listed values were extracted for thermally excited magnon transport, except for that of $\alpha$-$Fe_2O_3$[10,11,36-39,51-53,55-57]. Remarkably, $CrCl_3$ demonstrates the longest $\lambda_m$ among any MI reported so far. The observation of both superfluid spin transport in the canted AFM state and conventional magnon diffusion with exceptionally long $\lambda_m$ in the fully aligned spin state underscores the strong potential of vdW AFIs such as $CrCl_3$ for next-generation, ultra-low-dissipation spintronic technologies.

In conclusion, We demonstrate field-tunable spin-superfluid transport in the vdW AFI $CrCl_3$ using a nonlocal architecture. Consistently with theoretical predictions, we find that in the canted AFM state, the nonlocal response exhibits the algebraic distance dependence expected for a phase-stiff supercurrent, while strong fields or elevated temperature quench the superfluid and restore conventional, exponentially decaying magnon diffusion. Importantly, using this nonlocal device geometry generated by thermal excitation, we report the longest pure spin current transport distance to date (91 μm) in any magnetic system, together with a record $\lambda_m$ of 14 ± 3 μm. These results highlight the promise of vdW magnetic materials as a versatile platform for studying novel quantum phenomena such as magnon condensation and pave the way toward the development of low-dissipation spintronic devices based on superfluid spin transport.

## References

1 Bardeen, J., Cooper, L. N. & Schrieffer, J. R. Theory of Superconductivity. *Phys. Rev.* **108**, 1175-1204 (1957).

2 Allen, J. F. & Misener, A. D. Flow of Liquid Helium II. *Nature* **141**, 75-75 (1938).

3 Kapitza, P. Viscosity of Liquid Helium below the λ-Point. *Nature* **141**, 74-74 (1938).

4 London, F. The λ-Phenomenon of Liquid Helium and the Bose-Einstein Degeneracy. *Nature* **141**, 643-644 (1938).

5 Andersson, N. & Comer, G. L. Probing Neutron-Star Superfluidity with Gravitational-Wave Data. *Phys. Rev. Lett.* **87**, 241101 (2001).

6 Hennigar, R. A., Mann, R. B. & Tjoa, E. Superfluid Black Holes. *Phys. Rev. Lett.* **118**, 021301 (2017).

7 Arute, F. *et al.* Quantum supremacy using a programmable superconducting processor. *Nature* **574**, 505-510 (2019).

8 Devoret, M. H. & Schoelkopf, R. J. Superconducting Circuits for Quantum Information: An Outlook. *Science* **339**, 1169-1174 (2013).

9 Sonin, E. B. Spin currents and spin superfluidity. *Adv. Phys.* **59**, 181-255 (2010).

10 Cornelissen, L. J., Liu, J., Duine, R. A., Youssef, J. B. & van Wees, B. J. Long-distance transport of magnon spin information in a magnetic insulator at room temperature. *Nat. Phys.* **11**, 1022-1026 (2015).

11 Lebrun, R. *et al.* Tunable long-distance spin transport in a crystalline antiferromagnetic iron oxide. *Nature* **561**, 222-225 (2018).

12 Wei, X. Y. *et al.* Giant magnon spin conductivity in ultrathin yttrium iron garnet films. *Nat. Mater.* **21**, 1352-1356 (2022).

13 Yuan, W. *et al.* Experimental signatures of spin superfluid ground state in canted antiferromagnet $Cr_2O_3$ via nonlocal spin transport. *Sci. Adv.* **4**, eaat1098.

14 Althammer, M. All-Electrical Magnon Transport Experiments in Magnetically Ordered Insulators. *Phys.Status Solidi RRL.* **15**, 2100130 (2021).

15 Wimmer, T. *et al.* Spin Transport in a Magnetic Insulator with Zero Effective Damping. *Phys. Rev. Lett.* **123**, 257201 (2019).

16 Takei, S. & Tserkovnyak, Y. Superfluid Spin Transport Through Easy-Plane Ferromagnetic Insulators. *Phys. Rev. Lett.* **112**, 227201 (2014).

17 Flebus, B., Bender, S. A., Tserkovnyak, Y. & Duine, R. A. Two-Fluid Theory for Spin Superfluidity in Magnetic Insulators. *Phys. Rev. Lett.* **116**, 117201 (2016).

18 Takei, S., Halperin, B. I., Yacoby, A. & Tserkovnyak, Y. Superfluid spin transport through antiferromagnetic insulators. *Phys. Rev. B* **90**, 094408 (2014).

19 Evers, M. & Nowak, U. Transport properties of spin superfluids: Comparing easy-plane ferromagnets and antiferromagnets. *Phys. Rev. B* **101**, 184415 (2020).

20 Gong, C. & Zhang, X. Two-dimensional magnetic crystals and emergent heterostructure devices. *Science* **363**, eaav4450 (2019).

21 Gibertini, M., Koperski, M., Morpurgo, A. F. & Novoselov, K. S. Magnetic 2D materials and heterostructures. *Nat. Nanotechnol.* **14**, 408-419 (2019).

22 Zhou, Y., Li, S., Liang, X. & Zhou, Y. Topological Spin Textures: Basic Physics and Devices. *Adv. Mater.* **37**, 2312935 (2025).
23 He, Q. L. *et al.* Tailoring exchange couplings in magnetic topological-insulator/antiferromagnet heterostructures. *Nat. Mater.* **16**, 94-100 (2017).
24 Klein, D. R. *et al.* Enhancement of interlayer exchange in an ultrathin two-dimensional magnet. *Nat. Phys.* **15**, 1255-1260 (2019).
25 Bedoya-Pinto, A. *et al.* Intrinsic 2D-XY ferromagnetism in a van der Waals monolayer. *Science* **374**, 616-620 (2021).
26 Wang, Q. H. *et al.* The Magnetic Genome of Two-Dimensional van der Waals Materials. *ACS Nano* **16**, 6960-7079 (2022).
27 Cai, X. *et al.* Atomically Thin $CrCl_3$: An In-Plane Layered Antiferromagnetic Insulator. *Nano Lett.* **19**, 3993-3998 (2019).
28 Kuhlow, B. Magnetic Ordering in $CrCl_3$ at the Phase Transition. *Physica Status Solidi (a)* **72**, 161-168 (1982).
29 McGuire, M. A. *et al.* Magnetic behavior and spin-lattice coupling in cleavable van der Waals layered $CrCl_3$ crystals. *Phys. Rev. Mater.* **1**, 014001 (2017).
30 Narath, A. & Davis, H. L. Spin-Wave Analysis of the Sublattice Magnetization Behavior of Antiferromagnetic and Ferromagnetic $CrCl_3$. *Phys. Rev.* **137**, A163-A178 (1965).
31 Lu, X., Fei, R., Zhu, L. & Yang, L. Meron-like topological spin defects in monolayer $CrCl_3$. *Nat. Commun.* **11**, 4724 (2020).
32 Augustin, M., Jenkins, S., Evans, R. F. L., Novoselov, K. S. & Santos, E. J. G. Properties and dynamics of meron topological spin textures in the two-dimensional magnet $CrCl_3$. *Nat. Commun.* **12**, 185 (2021).
33 Wang, Z. *et al.* Determining the phase diagram of atomically thin layered antiferromagnet $CrCl_3$. *Nat. Nanotechnol.* **14**, 1116-1122 (2019).
34 Glamazda, A., Lemmens, P., Do, S. H., Kwon, Y. S. & Choi, K. Y. Relation between Kitaev magnetism and structure in $\alpha$-$RuCl_3$. *Phys. Rev. B* **95**, 174429 (2017).
35 Wang, J. *et al.* Physical Vapor Transport Growth of Antiferromagnetic $CrCl_3$ Flakes Down to Monolayer Thickness. *Adv. Sci.* **10**, 2203548 (2023).
36 Xing, W. *et al.* Magnon Transport in Quasi-Two-Dimensional van der Waals Antiferromagnets. *Phys. Rev. X* **9**, 011026 (2019).
37 de Wal, D. K. *et al.* Long-distance magnon transport in the van der Waals antiferromagnet $CrPS_4$. *Phys. Rev. B* **107**, L180403 (2023).
38 Feringa, F., Vink, J. M. & van Wees, B. J. Spin Nernst magnetoresistance for magnetization study of $FePS_3$. *Phys. Rev. B* **107**, 094428 (2023).
39 Yuan, P. *et al.* Unconventional Magnon Transport in Antiferromagnet $NiPS_3$ Induced by an Anisotropic Spin-Flop Transition. *Nano Lett.* **25**, 5350-5357 (2025).
40 Goennenwein, S. T. B. *et al.* Non-local magnetoresistance in YIG/Pt nanostructures. *Appl. Phys. Lett.* **107**, 172405 (2015).
41 Gomez-Perez, J. M., Vélez, S., Hueso, L. E. & Casanova, F. Differences in the magnon diffusion length for electrically and thermally driven magnon currents in $Y_3Fe_5O_{12}$. *Phys. Rev. B* **101**, 184420 (2020).
42 Yan, W. *et al.* Large room temperature spin-to-charge conversion signals in a few-layer graphene/Pt lateral heterostructure. *Nat. Commun.* **8**, 661 (2017).
43 Sagasta, E. *et al.* Tuning the spin Hall effect of Pt from the moderately dirty to the superclean regime. *Phys. Rev. B* **94**, 060412 (2016).
44 Cornelissen, L. J., Peters, K. J. H., Bauer, G. E. W., Duine, R. A. & van Wees, B. J. Magnon spin transport driven by the magnon chemical potential in a magnetic insulator. *Phys. Rev. B* **94**, 014412 (2016).
45 Uchida, K. *et al.* Spin Seebeck insulator. *Nat. Mater.* **9**, 894-897 (2010).
46 Flebus, B., Upadhyaya, P., Duine, R. A. & Tserkovnyak, Y. Local thermomagnonic torques in two-fluid spin dynamics. *Phys. Rev. B* **94**, 214428 (2016).
47 Tserkovnyak, Y., Bender, S. A., Duine, R. A. & Flebus, B. Bose-Einstein condensation of magnons pumped by the bulk spin Seebeck effect. *Phys. Rev. B* **93**, 100402 (2016).

48 Giles, B. L. *et al.* Thermally driven long-range magnon spin currents in yttrium iron garnet due to intrinsic spin Seebeck effect. *Phys. Rev. B* **96**, 180412 (2017).

49 Shan, J. *et al.* Criteria for accurate determination of the magnon relaxation length from the nonlocal spin Seebeck effect. *Phys. Rev. B* **96**, 184427 (2017).

50 Giles, B. L., Yang, Z., Jamison, J. S. & Myers, R. C. Long-range pure magnon spin diffusion observed in a nonlocal spin-Seebeck geometry. *Phys. Rev. B* **92**, 224415 (2015).

51 Aguilar-Pujol, M. X. *et al.* Magnon currents excited by the spin Seebeck effect in ferromagnetic EuS thin films. *Phys. Rev. B* **108**, 224420 (2023).

52 Muduli, P. *et al.* Local and nonlocal spin Seebeck effect in lateral Pt–$Cr_2O_3$–Pt devices at low temperatures. *APL Mater.* **9**, 021122 (2021).

53 Qi, S. *et al.* Giant electrically tunable magnon transport anisotropy in a van der Waals antiferromagnetic insulator. *Nat. Commun.* **14**, 2526 (2023).

54 Wimmer, T. *et al*. Observation of antiferromagnetic magnon pseudospin dynamics and the Hanle effect. *Phys. Rev. Lett.* **125**, 247204 (2020).

55 Venkat, G. *et al.* Magnon diffusion lengths in bulk and thin film $Fe_3O_4$ for spin Seebeck applications. *Phys. Rev. Mater.* **4**, 075402 (2020).

56 Guo, C. Y. *et al.* Magnon valves based on YIG/NiO/YIG all-insulating magnon junctions. *Phys. Rev. B* **98**, 134426 (2018).

57 Gückelhorn, J. *et al.* Quantitative comparison of magnon transport experiments in three-terminal YIG/Pt nanostructures acquired via dc and ac detection techniques. *Appl. Phys. Lett.* **117** (2020).

# Methods

## Sample preparation

High-quality $CrCl_3$ crystals (>99.99%) were purchased from HQ Graphene. For the device fabrication, few-layer flakes of $CrCl_3$ were first exfoliated and transferred onto a cleaned $Si/SiO_x$ (300 nm) substrate in atmospheric conditions. A series of $CrCl_3$ devices were prepared and explored in this work with varying thicknesses between 8 and 60 nm. The lateral devices were fabricated by e-beam lithography with spin-coating a positive resist of poly(methyl methacrylate) (PMMA) to pattern the mask. Then, the deposition of Pt was done in a high-vacuum magnetron sputtering system ($3\times10^{-8}$ torr), followed by a lift-off process in acetone. The size of the Pt contact wires on $CrCl_3$ flakes is 300 nm width and over 20 μm length depending on the flake size, and the thickness is always fixed at 5 nm. After the fabrication, the samples are kept inside a $N_2$-filled box.

## Characterization

The thickness of $CrCl_3$ flake was calibrated by atomic force microscopy (Agilent 5500 SPM). The Raman spectra of bulk and thin flakes of $CrCl_3$ were carried out using Renishaw® inVia Qontor micro-Raman instrument with a 100× objective and an excitation wavelength of 532 nm. For measuring the magnon transport, the prepared samples were loaded into a physical property measurement system (PPMS) by Quantum Design. A current was applied through the Pt wire by a Keithly 6221 and a nonlocal voltage signal ($V_{NL}$) was detected at the Pt detector by a Keithley 2182 nanovoltmeter. The injected current (*I*) was set in a delta mode ($I+, I-$) so we can extract the odd (linear in the leading order) and even (quadratic in the leading order) components of the nonlocal voltage signal ($V_{NL}$) from the equation: $V_{NL}^{odd} = \frac{V_{NL}(I+)-V_{NL}(I-)}{2}$ and $V_{NL}^{even} = \frac{V_{NL}(I+)+V_{NL}(I-)}{2}$. These correspond to the first-harmonic ($V_{NL}^{1\omega} \equiv V_{NL}^{odd}$) and second-haromonic ($V_{NL}^{2\omega} \equiv V_{NL}^{even}/2$) components of the ac lock-in techniques, respectively[57]. An external magnetic field was applied along the device plane.

# Acknowledgements

We acknowledge financial support from the Spanish MICIU/AEI/10.13039/501100011033 (CEX2020-001038-M), MICIU/AEI and ERDF/EU (PID2021-122511OB-I00) and by the European Union's Horizon 2020 Research and Innovation Programme (Grant Agreement No. 964396-SINFONIA). P.Y. and X.G. acknowledge support from the Spanish MICIU and European Union NextGenerationEU/PRTR (grant Nos. FJ2020-044666-I, JDC2022-049712-I, and RYC2021-031705-I, respectively). B.F. was supported by the U.S. Department of Energy, Office of Science, Basic Energy Sciences, under Grant No. DE- SC0024090.

# Author contributions

P.Y., F.C. and L.E.H conceived the study. X.G. performed flake exfoliation and characterization using Raman spectroscopy and atomic force microscopy. P.Y. carried out device fabrication and characterization using PPMS. B.F. and V.F. developed and performed the theoretical model. All authors discussed the data and contributed to writing the manuscript. F.C. and L.E.H. supervised the entire project.

# Additional information

Correspondence and requests for materials should be addressed to P.Y. (y.peisen@nanogune.eu), F.C. (f.casanova@nanogune.eu), and L.E.H. (l.hueso@nanogune.eu).

Supplementary Materials for

# Superfluid Spin Transport in the Van der Waals Antiferromagnet $CrCl_3$

Peisen Yuan[1,2,5*], Xiaomin Guo[1,5], Vincent Flynn[3], Benedetta Flebus[3], Fèlix Casanova[1,4*], and Luis E. Hueso[1,4*]

*[1]CIC nanoGUNE BRTA, 20018 Donostia-San Sebastian, Basque Country, Spain*

*[2]State Key Laboratory of Flexible Electronics (LoFE) & Institute of Flexible Electronics (IFE), Xiamen University, Xiamen 361102, China*

*[3]Department of Physics, Boston College, Chestnut Hill, Massachusetts 02467, United States*

*[4]IKERBASQUE, Basque Foundation for Science, 48009 Bilbao, Basque Country, Spain*

*[5]These authors contributed equally to this work.*

[*]Email: y.peisen@nanogune.eu, f.casanova@nanogune.eu, l.hueso@nanogune.eu

## Section 1. Theoretical model for spin superfluidity in 2D $CrCl_3$

While spin superfluidity is dynamically generated through the injection of spin angular momentum by thermal magnons, its variation with field is governed by equilibrium energetics; we therefore start from the free-energy functional of an undriven, nondissipative AFM bilayer:

$$F = J\boldsymbol{M_1} \cdot \boldsymbol{M_2} + \sum_{j=1}^{2} \left( \frac{A}{2} \left| \nabla \boldsymbol{M}_j \right|^2 + K\left(M_j^z\right)^2 - BM_j^x \right) \tag{1}$$

where $\boldsymbol{M_j}$ is the magnetization within the $j$'th layer, $J > 0$ is the interlayer antiferromagnetic exchange, $\mathrm{A} > 0$ the intralayer ferromagnetic stiffness, respectively, $K > 0$ is the easy-plane anisotropy strength, and $B$ is the applied magnetic field along the transport direction, chosen here to be along the $x$-axis.

In absence of the easy-plane anisotropy, i.e., $K = 0$, Eq. (1) is invariant under continuous rotations about the $x$-axis. For weak applied fields, the uniform ground state instead selects a particular orientation, spontaneously breaking this symmetry. To see this explicitly, we introduce the net magnetization $m$ and Néel vector $\boldsymbol{n}$ as

$$\boldsymbol{m} = \frac{\boldsymbol{M_1}+\boldsymbol{M_2}}{2M_s}, \qquad \boldsymbol{n} = \frac{\boldsymbol{M_1}-\boldsymbol{M_2}}{2M_s}, \tag{2}$$

with $M_s$ the saturation magnetization. For a spatially homogeneous state $\boldsymbol{m} = (m, 0,0)$ the bilayer free energy (with $K = 0$) Eq. (1) becomes

$$F = 2M_s(JM_s m^2 - Bm) + F_0 \tag{3}$$

where and $F_0$ is a constant. For $B < B_{sat} = 2JM_s$, Eq. (3) is minimized by a canted state with $m = m_{eq} = \frac{B}{B_{sat}}$, while for $B > B_{sat}$, the magnetization of each layer fully polarizes along the field, i.e., $m_{eq} = 1$ (see Figure 1d in the main text). In the fully polarized phase, the non-linear constraint $|\boldsymbol{m}|^2 + |\boldsymbol{n}|^2 = 1$ forces the Néel vector to vanish, i.e., $|\boldsymbol{n}| = 0$. On the other hand, in the canted regime, the Néel vector has a finite magnitude and, due to the nonlinear constraint $\boldsymbol{m} \cdot \boldsymbol{n} = 0$, is constrained to lie within the $yz$-plane. The free energy is, however, invariant under continuous rotations of the Néel vector within the $yz$-plane, signalling the emergence of a Goldstone mode, i.e. *the spin superfluid*[1,2].

In the presence of finite anisotropy, this U(1) symmetry is explicitly broken, and the Néel azimuthal angle $\varphi$ is pinned to $\varphi = 0$ or $\varphi = \pi$, where $\varphi$ is the rotation angle of $\mathbf{n}$ in the $yz$-plane. Any small rotation away from these values cost an energy $\propto K$ - however, when the anisotropy is sufficiently weak, thermal-magnon spin injection into the low-energy mode can overcome the easy-plane pinning and restore phase-coherent U(1) precession in the driven state[3].

To capture this interplay, we model the magnetization dynamics of the AFM bilayer in the canted state through a set of coupled Landau-Lifshitz-Gilbert equations, i.e.,

$$\frac{\partial \boldsymbol{M_j}}{\partial t} = -\gamma \boldsymbol{M_j} \times \boldsymbol{H_j} + \frac{\alpha}{M_s} \boldsymbol{M_j} \times \frac{\partial \boldsymbol{M_j}}{\partial t} \tag{4}$$

where $\boldsymbol{H_j} = -\frac{\delta F}{\delta \boldsymbol{M_j}}$ is the effective (i.e., Landau-Lifshitz) field and $\alpha > 0$ the Gilbert damping parameter. Neglecting contributions arising from spatial inhomogeneity of $m^x$, Eq. (4) can be rewritten as

$$\frac{\partial m^x}{\partial t} + \gamma \boldsymbol{\nabla} \cdot \boldsymbol{I} = \gamma K M_s (1-(m^x)^2) sin\,(2\varphi)\; + \alpha(1-(m^x)^2)\frac{\partial \varphi}{\partial t} \tag{5a}$$

$$\frac{\partial \varphi}{\partial t} = \gamma(B - B_{\text{sat}} m^x + 2KM_s m^x(\varphi)\; + AM_s m^x |\nabla\varphi|^2) - \frac{\alpha}{1-(m^x)^2}\frac{\partial m^x}{\partial t} \tag{5b}$$

where $\boldsymbol{I} = AM_s(1-(m^x)^2)\boldsymbol{\nabla}\varphi$ is the spin supercurrent. In the absence of both anisotropy and damping, the magnetization component $m^x$ and the Néel-angle $\varphi$ form a canonical pair and evolve according to Hamilton's equations. Equation (5a) then becomes a continuity equation for $m^x$, where the spin current is carried by spatial gradients of $\varphi$, and the AFM exchange $J$ plays the same role of the easy-plane anisotropy in the ferromagnetic spin-superfluid states[1,4,5].

Eqs. (5a)-(5b) admit stationary solutions with a uniform phase gradient -a spin supercurrent that decays algebraically over long distances, in stark contrast to the exponential attenuation of spin signals carried by incoherent, thermal magnons. To derive these solutions, we consider a quasi-1D geometry and focus on magnetization textures that are slowly varying in space. Specifically, we assume $|\boldsymbol{\nabla}\varphi|^2 \ll B_{sat}/AM_s$, which allow us to neglect the $|\boldsymbol{\nabla}\varphi|^2$ term in Eq. (5b). The heat flux across the Pt-$CrCl_3$ interface ($x = 0$ ) drives a non-equilibrium magnon population in the AFM sample, which can be quantified in terms of a local non-equilibrium magnetization $m' = m^x - m_{eq} \ll 1$. We assume that the latter is converted into a coherent superfluid spin flow over a distance negligible compared to the sample length $d$, allowing us to neglect two-fluid (thermal-magnon) backflow. Taking $\boldsymbol{I}(x=0) = I_0\hat{\boldsymbol{x}}$ to be the current injected into the superfluid mode, assuming a sufficiently weak anisotropy (see below for a discussion on this assumption), and setting $\partial m^x/\partial t = 0$ in Eq. (5), we obtain the spin supercurrent

$$\boldsymbol{I}(x) = I_0\left(1 - \frac{x}{d+L_\alpha}\right) \tag{6}$$

where $L_\alpha = v_d B_{sat}/[\alpha(B_{sat}^2 - B^2)]$ is a characteristic decay length and $v_d$ is a phenomenological constant relating the spin current at the detector to the nonequilibrium spin accumulation[6]: $I(d) = v_d m'$, where

$$m' = \frac{L_\alpha}{v_d}\frac{I_0}{d+L_\alpha}\,. \tag{7}$$

Note that these results are derived assuming $|m'| \ll 1$. Since $m'$ increases with $B/B_{sat}$ according to Eq. (7), we only expect these results to hold in the weak-field regime wherein $L_\alpha$ is approximately independent of $B$. The phenomenological constant $v_d$ can be related to the interfacial spin mixing conductance $g^{\uparrow\downarrow}$ by employing the spin-current boundary conditions derived in Ref. (2). Namely, $v_d = g^{\uparrow\downarrow}\hbar\gamma B_{sat}/4\pi M_s$.

Equation (6) shows that the spin current detected at the interface $x = d$ through its conversion into a voltage via the inverse Hall effect decays algebraically with $d$. In the weak-field regime $B \ll B_{sat}$, the characteristic length scale $L_\alpha$ can be shown to be largely insensitive to $B$ :

$$L_\alpha = \frac{g^{\uparrow\downarrow}}{4\pi M_s \alpha}\hbar\gamma\,\frac{1}{1-(B/B_{sat})^2} \tag{8}$$

This is consistent with the nearly constant dependence of $L_\alpha$ on $B$ demonstrated experimentally in Figure 3d.

For a rough microscopic estimate of $L_\alpha$, $g^{\uparrow\downarrow}$ can be extracted from our spin Hall magnetoresistance (SMR) measurements (see Section 2). According to previous reports, $g^{\uparrow\downarrow}$ can be extracted from the longitunidual amplitude of SMR ($\Delta\rho_L/\rho_0$) by the following equation:[6]

$$\frac{\Delta\rho_L}{\rho_0} = \theta_{SH}^2 \frac{\lambda_s}{d_N}\left[\frac{2g^{\uparrow\downarrow}tanh^2(\frac{d_N}{2\lambda_s})}{\sigma+2\lambda_s g^{\uparrow\downarrow}coth\frac{d_N}{\lambda_s}}\right] \tag{9}$$

where $\rho_L$, $\theta_{SH}$, $\lambda_S$, $d_N$, $\sigma$ (=1/$\rho_0$) are the resistivity, spin Hall angle, spin diffusion length, thickness and conductivity of Pt layer, respectively. From our SMR results, we set $\rho_0$=71 μΩ•cm, $d_N = 5\,nm$, $\Delta\rho_L/\rho_0$=3.5×10$^{-5}$. Additionally, both $\theta_{SH}$ and $\lambda_S$ of Pt can be obtained from our previous work due to their special resistivity (ρ) dependence and are set as $\theta_{SH} \sim 0.1$ and $\lambda_S \sim 0.7\,nm$, accordingly.[7] As a result, $g^{\uparrow\downarrow}$ is calculated from Eq. (9) to be around 2.61×10$^{13}$ Ω$^{-1}$•m$^{-2}$. For evaluating $\alpha$ in a thin layer $CrCl_3$, in the absence of direct measurements for our specific Pt|$CrCl_3$ heterostructures, we first obtain a rough estimate by inserting a value from a bulk crystal of $CrCl_3$ under ferromagnetic resonance mode: $\alpha$~$2.0 \times 10^{-3}$.[8] Evaluating the expression in Eq.(1) at B=0.3 T yields $L_\alpha$ is below 0.01 μm, smaller than the $L_\alpha$~ 5 μm extracted from our nonlocal data.

This discrepancy reflects a well-known limitation of the LLG phenomenology when it is used to model longitudinal spin relaxation: the single Gilbert parameter $\alpha$ lumps together spin-conserving and spin-nonconserving channels, whereas the hydrodynamic quantity $1/T_1$ entering the spin-superfluid theory represents only spin-nonconserving processes and is controlled by spin-orbit coupling.[9] In our insulating $CrCl_3$ samples, the microscopic anisotropies imply spin-orbit energies that are orders of magnitude smaller than the exchange scale, so the true $T_1$ is expected to be much longer than what a naive LLG estimate would suggest. We therefore view Eq. (1) in the main text as a phenomenological parametrization and interpret the fitted $L_\alpha$ as an effective decay length that encapsulates all relevant relaxation channels in our Pt/$CrCl_3$ heterostructures. Under this assumption, the experimentally extracted $L_\alpha$ is fully consistent with the exceptionally weak spin-orbit coupling of $CrCl_3$ and with the long spin-relaxation lengths reported in related van der Waals magnets. Moreover, the hallmark transport signature of a spin-superfluid state is an algebraic decay of the spin current with distance. This behavior arises generically in any superfluid system subjected to longitudinal dissipation; the specific form of the dissipation only sets the characteristic decay length.

It is important to remark that applicability of our spin-superfluid theory is confined to regimes where the thermal-magnon density remains low. At higher temperatures, magnon-magnon scattering rates and Gilbert damping grow rapidly, destroying the long-range phase coherence necessary for a superfluid spin current. Incoherent thermal magnons become then the dominant spin carriers, and the algebraic decay of the supercurrent gives way to an exponential attenuation[4].

Similarly, beyond the weak-field regime, the spins align resulting in an effective single FM degree of freedom. In this "spin-polarized" regime, the U(1) phase degree of freedom is lost, and spin transport is once again carried by diffusive magnons with an exponential decay

$\boldsymbol{I}(x) \propto \boldsymbol{\nabla} m^x \sim e^{-x/\lambda_m}$, where $\lambda_m$ is the magnon diffusion length.[9]

Finally, we review the assumptions underlying the derivation of Eq (6) within the weak-field and low-temperature regime. With the solution in-hand, we find that the small gradient assumption on $\varphi$ is equivalent to requiring the injected spin current to remain below a critical value, specifically,

$$I_0^2 \ll AM_s B_{sat} \tag{10}$$

At the same time, the weak anisotropy $K$ introduces a source term into the equation of motion for $\boldsymbol{m}^x$ that breaks spin current conservation by penalizing phase gradients. However, a sufficiently large injection of angular momentum can overcome both this anisotropy and intrinsic damping, enabling the system to reach the threshold for condensation.[4] Mathematically, this condition translates[1]

$$AM_s^2 K \ll I_0^2. \tag{11}$$

Given an anisotropy field of 300 mT and known material properties of $CrCl_3$,[10,11] we estimate the lower bound of spin-current to be $I_0M_s \sim 4.7\times10^{-4}$ J m$^{-2}$. Consistency between these two conditions Eq. (10) and (11) translates to $M_s K \ll B_{sat}$, which, in words, translates to the requirement that the in-plane anisotropy field is significantly weaker than the saturation field (which, itself, is proportional to the interlayer AFM exchange field) - which is well-known to be the case for this material.

The absence of a first-harmonic signal and the linear decay in the spin current observed in our main devices is consistent with the hypothesis that the spin superfluidity is mainly generated via the conversion of thermally injected magnons over a finite distance $\xi$ from the injection interface,[5] with $\xi \ll d$, where $d$ is the sample length. This conversion occurs over a characteristic healing length, determined by the interplay between spin stiffness and dissipation. Depending on the dominant dissipation mechanism -whether magnon-magnon or magnon-defect (or phonon) scattering- the healing length can be estimated as either the coherence length or the magnon mean free path. Of the two, the mean free path is generally longer.[9] In similar antiferromagnets (e.g. $CrI_3$), the magnon mean free path is found to be order of 100 nm,[12] i.e., much shorter than the sample length; by analogy, we expect $\xi \ll d$ to hold in our devices as well.

It should be noted that we model the sample as a continuous system because the lattice constant (on the order of a few angstroms between Cr atoms) is many orders of magnitude smaller than the relevant length scales in our experiment, including both the sample size (µm scale) and the sample thickness (10-30 nm). Under these conditions, a continuum description is fully accepted for modelling spin transport in such systems. Furthermore, boundary "bounce-back" effects that can occur for conventional thermally driven magnons are not applicable to superfluid spin transport. This distinction arises because incoherent magnon transport and superfluid spin transport are fundamentally different physical mechanisms.

1) Transport by incoherent magnons (the "usual" magnons, whose contribution typically decays exponentially with distance). These can be viewed as a collection of quasi-particles that carry angular momentum individually, propagate independently of each other and can scatter in the sample. In a finite geometry, such magnons can indeed reflect from sample boundaries and interfaces.

2) Transport by a spin superfluid,[1] which is a hydrodynamic, phase-coherent mode of the order parameter. The associated spin current does not emerge from a collection of individual, propagating magnons; rather, it is carried by a coherent twist (a phase gradient) of the order parameter.

Using an analogy, superfluid spin transport is like transmitting angular momentum by twisting a rod, normal spin-wave transport is more akin to throwing many little spinning balls.[1] The balls can scatter and bounce at interfaces, depleting the angular momentum that reaches the detector. For the twisted rod, “bounce-back” is not the right picture; boundaries mainly impose constraints on the twist (and can relax it via slip events), rather than reflecting carriers.

**Section 2. Local spin Hall magnetoresistance (SMR) measurement**

The spin injection at the Pt/$CrCl_3$ interface is verified by measuring the SMR of a Pt strip deposited on a $CrCl_3$ flake (see details in Supplementary Figure 2). For simplicity, the measurements are performed at 4 T and below $T_N$. At this field, the spins in $CrCl_3$ are fully aligned along the external magnetic field since the saturation field $B_{sat}$ is around 2 T. The normalized longitudinal angle-dependent magnetoresistance (ADMR) exhibits a maximum (minimum) value when $B$ is parallel (perpendicular) to the current $I$ (Supplementary Figure 3a). It follows a $sin^2\theta$ dependence, with $\theta = 0°$ defined when $B$ is perpendicular to $I$. Such behavior is consistent with the interplay of the SHE and ISHE at the Pt/$CrCl_3$ interface, a mechanism extensively studied in other magnetic insulator (MI)/heavy metal systems, such as the YIG/Pt heterostructure.[13] Furthermore, the ADMR signal is only visible below $T_N$ (Supplementary Figure 3b), directly reflecting its correspondence with the magnetic ordering in $CrCl_3$. In contrast, the normalized field-dependent magnetoresistance (FDMR) displays an anomalous enhancement when increasing either the current or the temperature -even above $T_N$ (up to 50 K, Supplementary Figures 3c and 3d). This behavior may originate from the formation of a disordered Pt/$CrCl_3$ interface during Pt deposition, which could induce extrinsic magnetic effects at the Pt/$CrCl_3$ interface.[14] However, this disordered interface primarily influences the first-harmonic signal, while the second-harmonic voltage ( $V_L^{2\omega}$ ), remains unaffected (Supplementary Figures 3e-3h). The normalized ADMR of $V_L^{2\omega}$ exhibits a clear -$cos\theta$ angular dependence, as expected when only the ISHE at the detector is considered for thermal excited magnons. Notably, this $V_L^{2\omega}$ persists up to $T_N$, confirming its origin in the magnetic ordering of $CrCl_3$. Meanwhile, the normalized FDMR of $V_L^{2\omega}$ shows the same trend under varying current and temperature conditions below $T_N$ (Supplementary Figure 3g and 3h), with a saturation field around 2 T that agrees well with the previous magnetotransport measurement of $CrCl_3$.[15,16].

**Section 3. First-harmonic nonlocal signal in $CrCl_3$ devices**

We note that no $V_{NL}^{1\omega}$ is detected in our $CrCl_3$ devices with thin flakes, even for short injector-detector separations down to $d$=1 μm (Supplementary Figure 7 and 8). This observation is not unique to $CrCl_3$. To date, among all reported Pt/2D magnetic-insulator nonlocal heterostructures,[17-22] $CrPS_4$ is the only two-dimensional magnetic insulator in which a $V_{NL}^{1\omega}$ signal has been experimentally observed.[18,22] Even in those studies, $V_{NL}^{1\omega}$ signals are extremely weak and accompanied by substantial background noise, and are not consistently observed across devices. For example, in thinner $CrPS_4$ flakes or in devices with longer $d$, no detectable $V_{NL}^{1\omega}$ signal was reported.[19] These results strongly suggest that electrically excited magnon injection in such Pt-based nonlocal geometries is generally inefficient in 2D magnetic insulators, with detectable $V_{NL}^{1\omega}$ signals limited to very thick magnetic layers and submicron channel lengths.

Consistent with this picture, in our $CrCl_3$ devices we detect a finite but weak $V_{NL}^{1\omega}$ signal only in very thick flakes (over 80 nm) and for short separations (below 1 μm) (Supplementary Figure 8-10). As shown in Supplementary Figure 9a and 9b, we present the angle-dependent $V_{NL}^{1\omega}$ and second-harmonic ($V_{NL}^{2\omega}$) nonlocal signals measured on very thick devices (>100 nm) at different temperatures. Notably, the $V_{NL}^{1\omega}$ signal becomes observable only at very low temperatures ($T \leq 5$ K) and remains extremely weak, with a background noise comparable to the signal itself at higher temperatures. Quantitatively, its amplitude at 2 K is about 10 times smaller than that of the $V_{NL}^{2\omega}$, indicating that electrically injected magnons are significantly less efficient than thermally generated magnons in our system. Furthermore, we measured the magnetic-field dependence of both nonlocal signals and found that the $V_{NL}^{1\omega}$ and $V_{NL}^{2\omega}$ responses exhibit similar line shapes (Supplementary Figure 9c and 9d), saturating at approximately 2 T, which corresponds to the saturation field ($B_{sat}$) of $CrCl_3$. We also confirmed that, as $d$ increases, the $V_{NL}^{1\omega}$ rapidly decays and becomes unmeasurable once the distance exceeds ~1 μm (Supplementary Figure 9e).

When the $CrCl_3$ thickness is reduced (while still remaining relatively thick, ~80 nm), a reliable $V_{NL}^{1\omega}$ can only be detected at 2 K (Supplementary Figure 10a) but its amplitude is around 3 times smaller than that of $V_{NL}^{2\omega}$ in above thick device (Supplementary Figure 9a). Upon increasing the temperature to 3 K, the signal becomes dominated by noise and falls below the instrumental detection limit, preventing the extraction of a clear angular dependence.

Overall, a measurable $V_{NL}^{1\omega}$ signal in $CrCl_3$/Pt nonlocal devices is observed only under very restricted conditions. It means that electrically injected magnon currents are significantly less efficient than thermally generated magnons in $CrCl_3$. The already weak $V_{NL}^{1\omega}$ signal is further suppressed at elevated temperatures, in thicker samples, and with increasing $d$, eventually approaching the detection limit of our measurement system. More importantly, in a magnetic system with finite anisotropy, a sufficiently large spin-angular-momentum injection is required to overcome anisotropy-induced pinning and restore coherent spin dynamics. The relatively weak magnon excitation generated by electrical injection is therefore insufficient to efficiently convert incoherent magnons into enough coherent spin-superfluid state capable of sustaining long-range transport. Under these conditions, the weak and readily weak $V_{NL}^{1\omega}$ response makes it difficult to reliably extract systematic long-range spin-transport characteristics. We therefore focus on thermally generated magnon currents, which provide a more efficient means of driving coherent spin dynamics and enable a systematic investigation of long-range superfluid spin transport in $CrCl_3$.

**Section 4. Temperature profile measurement in a nonlocal $CrCl_3$ device**

To unambiguously attribute the observed long-range nonlocal signal to superfluid spin transport, we carefully measured the temperature profile along the $CrCl_3$ channel using the experimental configuration illustrated in Supplementary Figure 12. A nonlocal device was fabricated with the same geometry as that used for the spin-transport measurements, except that each Pt strip was configured for independent four-probe resistance measurements. In this configuration, the Pt electrodes serve not only as spin injectors and Joule-heating sources but also as highly sensitive local thermometers, since the electrical resistance of Pt exhibits a well-defined dependence on temperature. When an injection current ($I_{NL}$) is applied to the injector electrode (considered as thermal generator), the generated Joule heat propagates through the $CrCl_3$ channel. By monitoring the resistance change of Pt electrodes (four-point measurement) located at different distances ($d$) from the injector, the temperature change at each position can be quantitatively determined, enabling direct visualization of the temperature profile along the transport channel.

We first calibrated the temperature dependence of the resistance of a Pt wire (considered as $Pt_{meter}$) using a four-probe configuration (Supplementary Figure 13a). At a base temperature of 2 K, the local current ($I_{Loc}$) applied to $Pt_{meter}$ was systematically varied to establish the relationship between the $I_{Loc}$ and its resistance (inset of Supplementary Figure 13a). As $I_{Loc}$ was reduced, the resistance gradually converged to a constant value. In this case, the Joule heating generated within $Pt_{meter}$ was very small and the effective local temperature of the detector ($T^1_{eff}$) was identical to the base temperature of 2 K (Supplementary Figure 13b). We then fixed $I_{Loc}$=1 μA, while a nonlocal injection current ($I_{NL}$) was applied to a neighboring Pt injector ($Pt_{inj}$). The Joule heat generated at $Pt_{inj}$ propagates through the $CrCl_3$ channel and modifies the temperature at $Pt_{meter}$, resulting in a second effective temperature ($T^2_{eff}$) (Supplementary Figure 13c). By comparing $T^1_{eff}$ (which is around 2 K) and $T^2_{eff}$, we extracted the local temperature change ($\Delta T$) at the detector position as a function of $I_{NL}$.

Supplementary Figure 13d shows the evolution of $\Delta T$ with $I_{NL}$ for an electrode separation of $d$=1 μm, while the corresponding results for other $d$ are presented in Supplementary Figure 14 using the same measurement. Supplementary Figure 14e summarizes the resulting temperature-change map as a function of both $d$ and $I_{NL}$. At a short electrode separation of $d$=1 μm, with the relatively small injection current of $I_{NL}$=10 μA, which is representative of the injection current used in the main text, generates certain Joule heating to produce a measurable temperature increase at the thermometer electrode. This result indicates that thermal gradients can indeed contribute to the interpretation of spin transport signals at very short distances. However, for $d \geqslant 2$ μm, the temperature increase induced by $I_{NL}$=10 μA becomes negligible. A measurable $\Delta T$ is observed only at substantially larger injection currents. We also measured the temperature profiles at different base temperatures and in two additional $CrCl_3$ devices with different thicknesses (Supplementary Figure 15), obtaining the same conclusions. Therefore, under the experimental conditions used for the spin transport measurements in the main text, the detected nonlocal spin signals at $d \geqslant 2$ μm and $I_{NL}$=10 μA are not accompanied by any temperature gradient, effectively excluding any contribution from thermally driven transport.

## Section 5. Discussion on the crossover between diffusive thermal magnons and spin-superfluid mode in short *d*

As discussed in Section 1, the spin-superfluid state in $CrCl_3$ is not an equilibrium phase that forms spontaneously upon cooling. Instead, it is dynamically generated through the injection of enough spin angular momentum carried by incoherent thermal magnons to overcome the anisotropy-induced pinning and restore the coherent spin dynamics. Consequently, at short transport distances, there exists an intrinsic conversion regime in which injected diffusive thermal magnons are converted into the spin-superfluid mode. This characteristic length scale corresponds to the healing length shown in Figure 1d of the main text.

For the short transport distance in the range of the healing length, we observed that the field-dependent $V_{NL}^{2\omega}$ measured in Device 1 exhibits a reproducible shoulder between approximately 0.5 and 1 T (Figure 2d). Note that this feature is reproducibly observed across multiple devices with different $CrCl_3$ thicknesses in short *d*, ruling out possible artifacts. A thin-layer odd-even spin-flop origin is also unlikely. The odd-even signatures reported for atomically thin $CrCl_3$ become negligible beyond approximately nine layers, whereas Device 1 is 20 nm thick.[16] Moreover, the shoulder emerges only below approximately 5 K (Supplementary Figure 20), well below the Néel temperature $T_N \sim 14$ K, rather than persisting throughout the ordered phase as expected for the reported spin-flop transition in odd-layer $CrCl_3$.

A phenomenological interpretation consistent with these observations is that the shoulder marks a crossover between coherent and diffusive contributions with different magnetic-field dependences:

$$V_{NL}^{2\omega}(B,d) = V_{super}(B,d) + V_{diff}(B,d) \tag{12}$$

where $V_{super}$ denotes the contribution associated with coherent spin-superfluid transport, while $V_{diff}$ denotes conventional diffusive transport by thermally excited magnons.

A shoulder can appear even if neither contribution is individually field independent, provided that their field derivatives partially compensate:

$$\frac{\partial V_{NL}^{2\omega}}{\partial B} = \frac{\partial V_{super}}{\partial B} + \frac{\partial V_{diff}}{\partial B} \approx 0 \tag{13}$$

This effect can be driven by the coexistence of transport channels: The conversion of thermally excited magnons into the spin-superfluid mode is most likely not completely efficient. Consequently, at short transport distances, such as $d = 1$ μm, which is much shorter than the magnon diffusion length, coherent spin-superfluid transport and conventional diffusive thermal magnon transport may coexist.

At low fields and temperatures, conversion into the coherent mode may be sufficiently efficient for the spin-superfluid contribution to remain substantial. As the field is increased, however, this contribution becomes progressively suppressed, i.e., $\partial V_{super}/\partial B < 0$ for $B \sim 0.7$ T. Meanwhile, the conventional diffusive contribution continues to increase within the experimentally relevant field range. The partial compensation between these two field dependences reduces the slope of the total voltage and produces the observed shoulder.

At higher fields, the coherent contribution is strongly suppressed, and conventional diffusive magnons dominate the short-distance signal. The field-induced imbalance between oppositely polarized magnon branches, together with progressive canting of the magnetic layers, can then increase the net spin carried by the diffusive magnon population, producing the renewed increase of the voltage before saturation. We present this as a physical interpretation consistent with the full data set, rather than as a unique identification based on the shoulder alone.

This interpretation is consistent with the disappearance of both the shoulder and the long-range superfluid-spin transport contribution above approximately 5 K (Supplementary Figure 17 and Figure 20). We also observed a shoulder feature in the first-harmonic nonlocal signal in our previous revised manuscript at a short $d$=900 nm (Supplementary Figure 9c and 10c), which may also indicates the crossover between the spin superfluidity and the normal electrically excited magnons, although the shoulder alone does not establish superfluid spin transport unambiguously; the principal evidence is the approximately algebraic distance dependence observed at low fields and temperatures, together with its systematic crossover to conventional exponential attenuation at higher fields or temperatures.

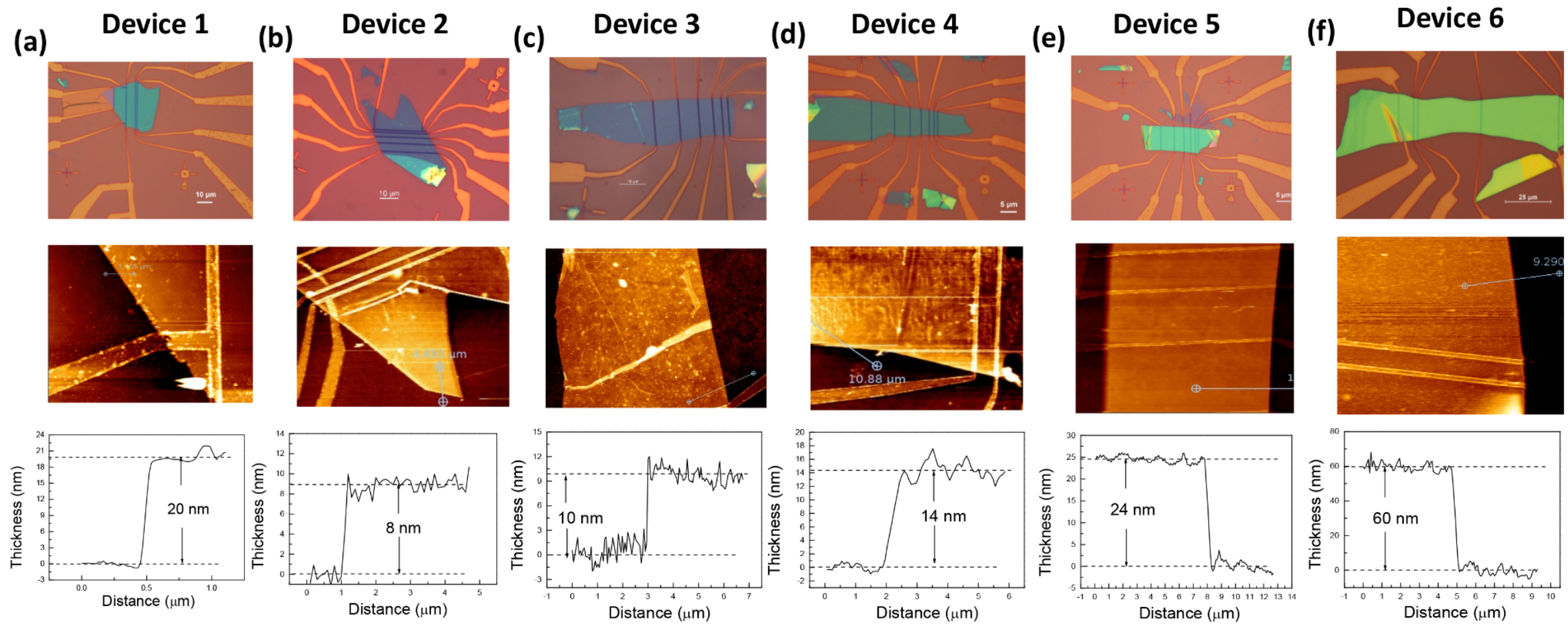

**Figure S1. Optical and atomic force microscopy (AFM) images of $CrCl_3$ devices with respective thicknesses** of (a) 20 nm (Device 1), (b) 8 nm (Device 2), (c) 10 nm (Device 3), (d) 14 nm (Device 4), (e) 24 nm (Device 5), and (f) 60 nm (Device 6). The line profiles shown below each AFM image correspond to the height variations measured along the solid lines indicated in the central AFM panels.

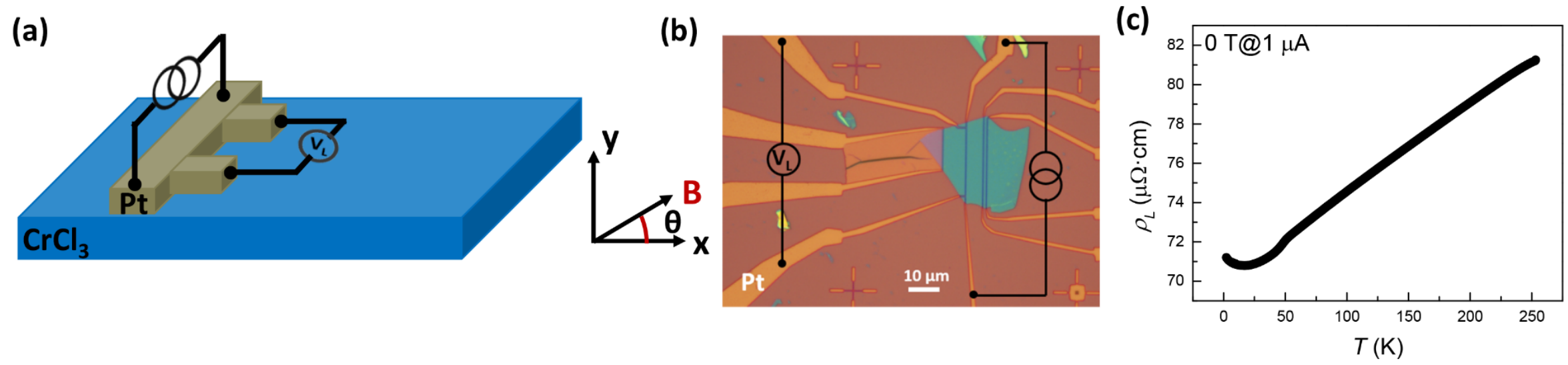

**Figure S2. Local voltage ( $V_L$ ) measurement setup for probing the spin Hall magnetoresistance.** (a) Schematic illustration of the $V_L$ measurement setup, employing a standard four-point configuration to measure the longitudinal resistance. (b) Optical image of the $CrCl_3$ device showing the $V_L$ measurement setup. (c) Temperature dependence of the longitudinal resistivity (ρ) of the Pt strip.

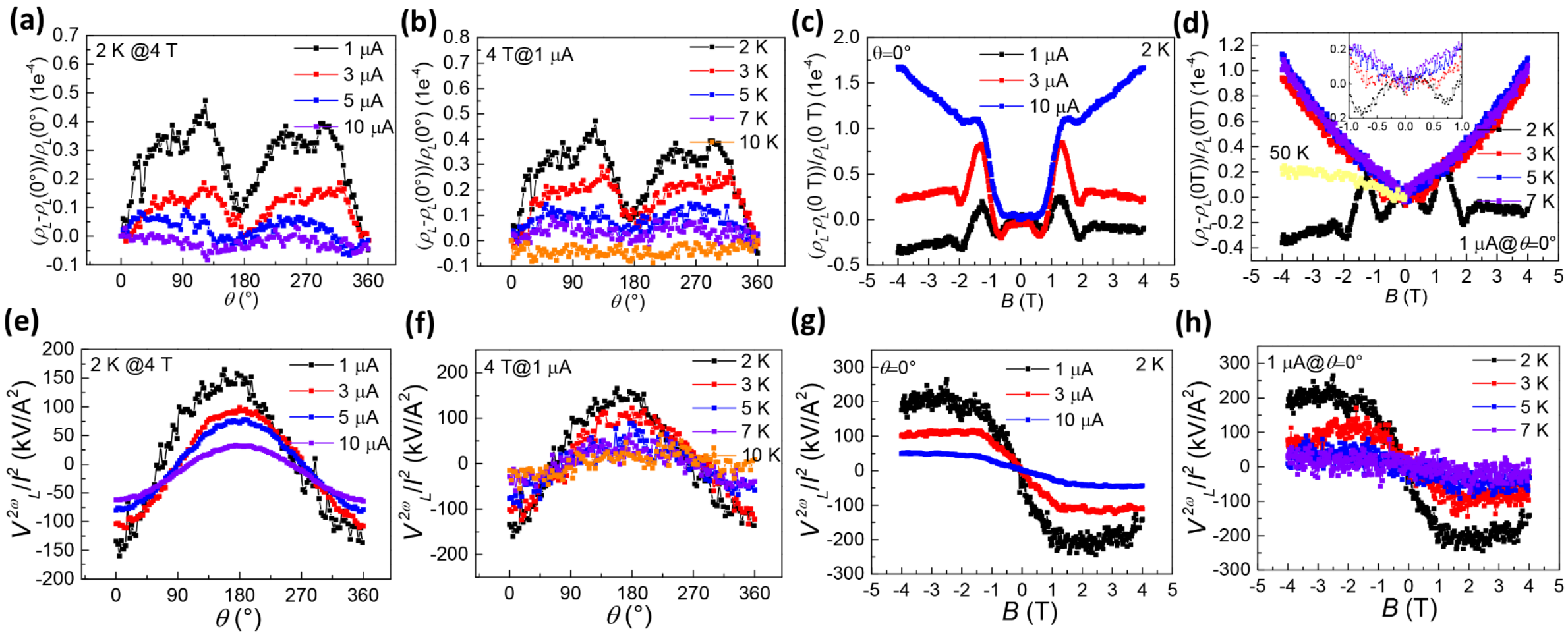

**Figure S3. Angle- and field-dependent SMR measurement.** (a, b) Longitudinal angle-dependent magnetoresistance (ADMR) measured at various injection currents (a) and temperatures (b), normalized to the corresponding base resistance of $\theta$=0°. (c, d) Longitudinal field-dependent magnetoresistance (FDMR) measured at various injection currents (c) and temperatures (d), normalized to the corresponding base resistance of $B$=0 T. (e, f) Angle-dependent second-harmonic local voltage signals ($V_L^{2\omega}$) obtained under varying injection currents (e) and temperatures (f). (g, h) Field-dependent second-harmonic local voltage signals ($V_L^{2\omega}$) obtained under varying injection currents (g) and temperatures (h). All $V_L^{2\omega}$ values are normalized by the square of the injection current.

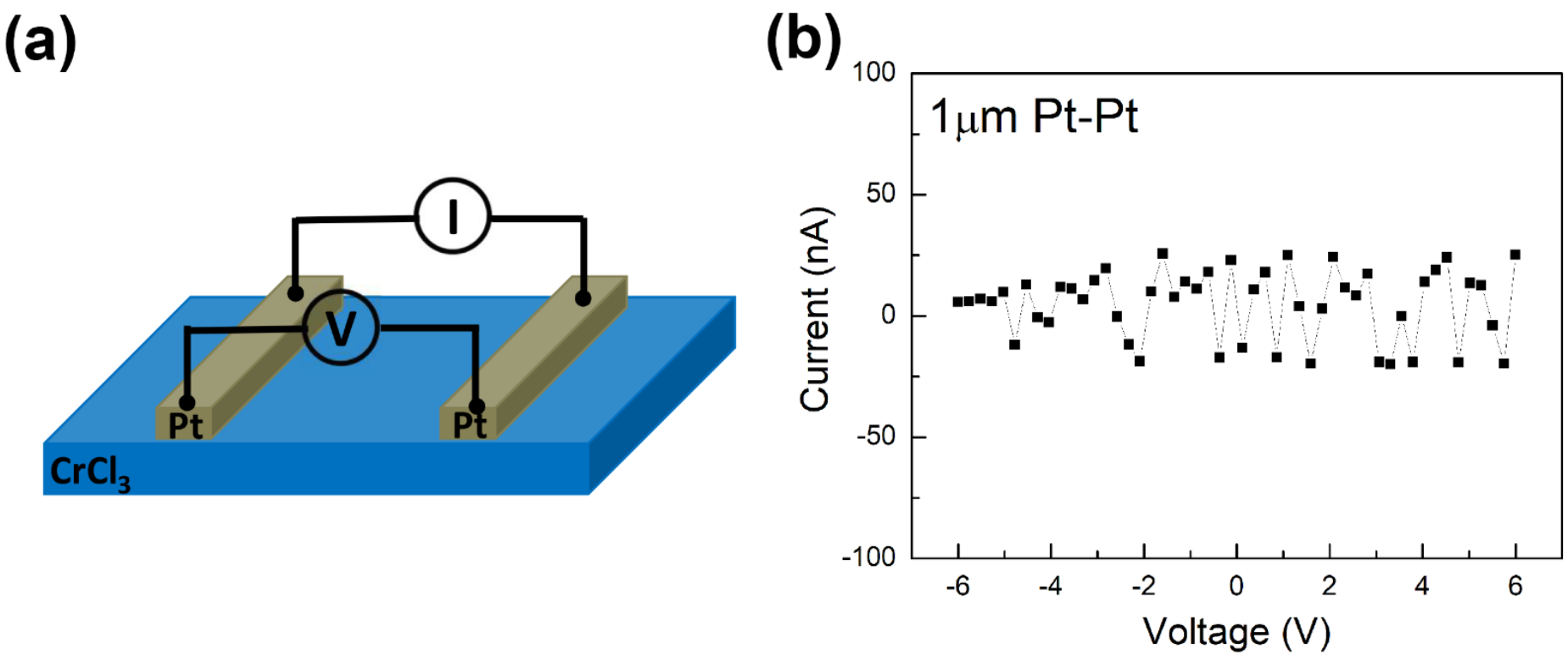

**Figure S4. Insulating properties of $CrCl_3$.** (a) Schematic illustration of the I–V measurement configuration across the $CrCl_3$ flake for Device 1. (b) Corresponding I–V characteristic of the $CrCl_3$ device measured using the setup shown in (a).

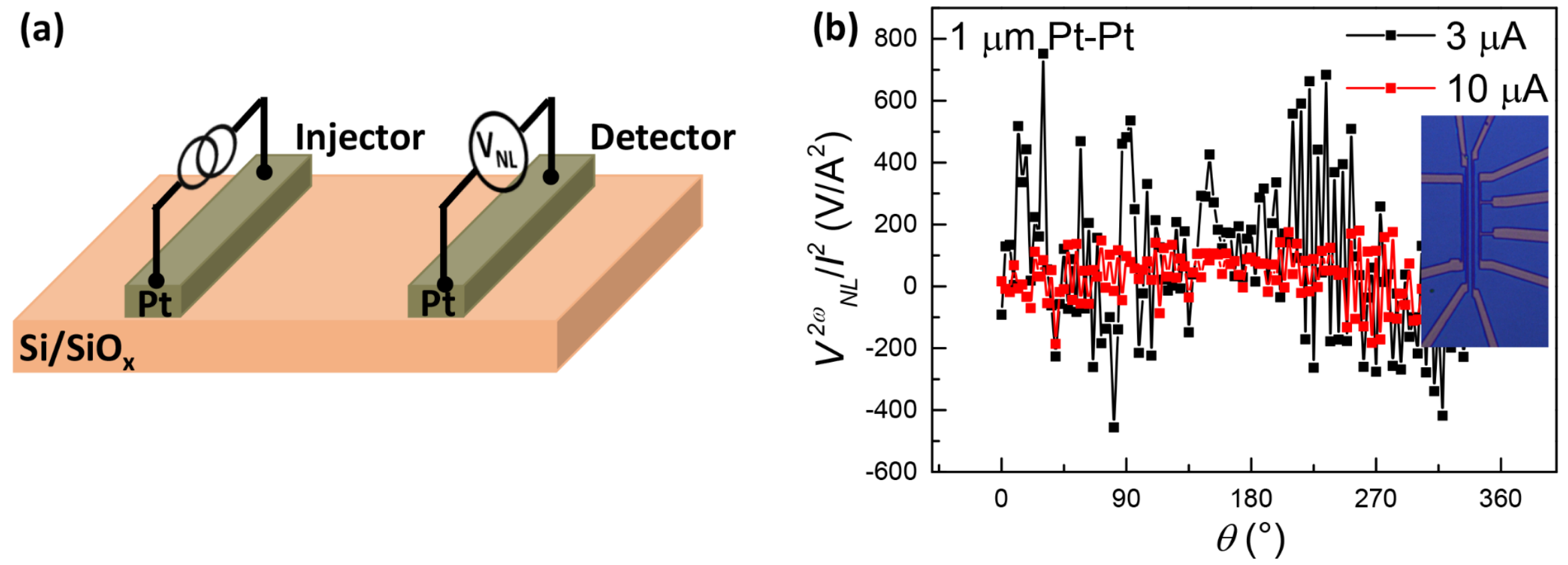

**Figure S5. Nonlocal voltage measurement of a control device comprising Pt strips on a $Si/SiO_x$ substrate.** (a) Schematic illustration of the nonlocal voltage measurement configuration for the control device. (b) Angle-dependent nonlocal voltage signals recorded at various injection currents. The inset shows an optical image of the control device.

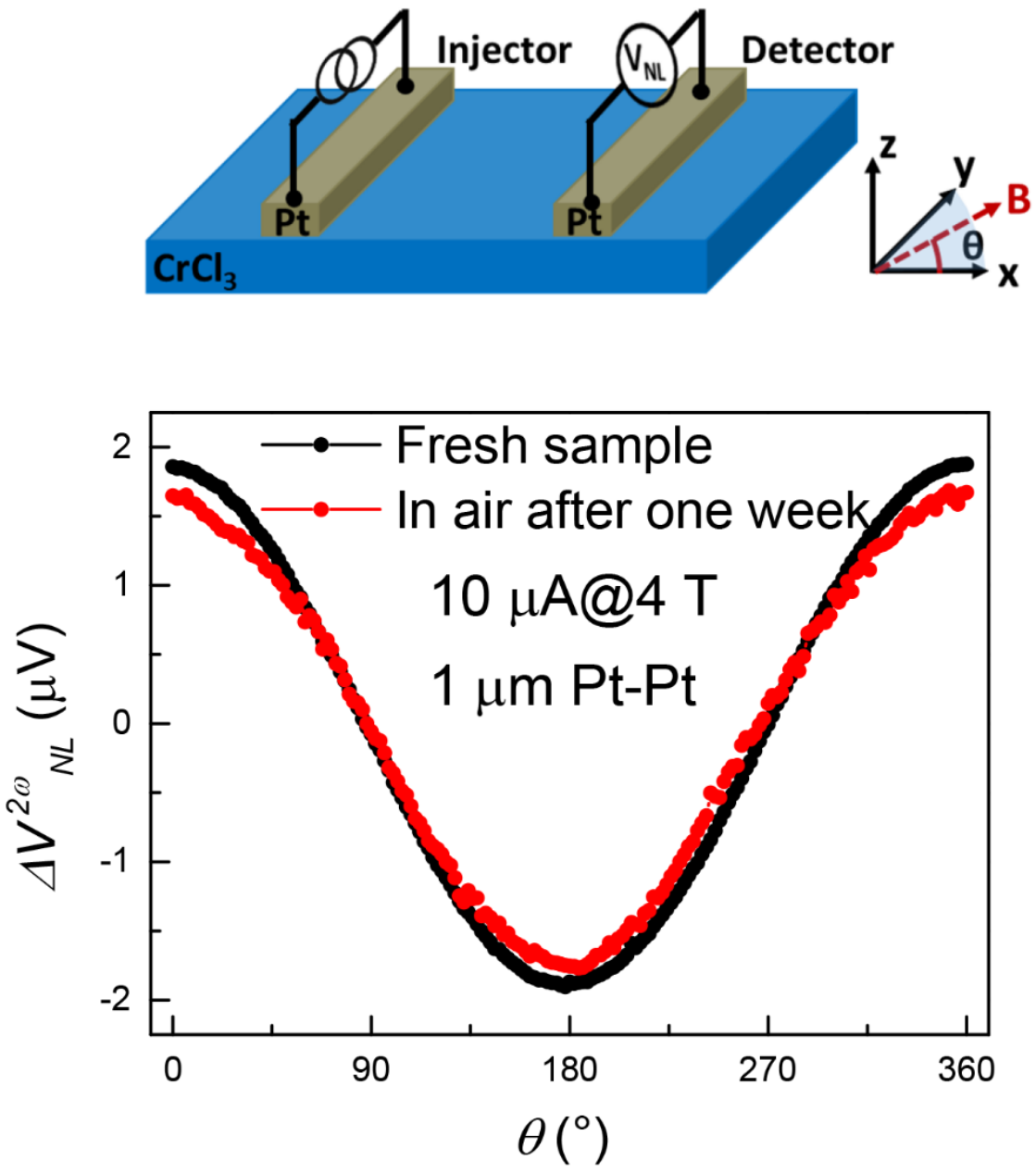

**Figure S6. Stability analysis of nonlocal magnon signals in a $CrCl_3$ device.** Top panel: Schematic diagram of the nonlocal voltage measurement configuration. Bottom panel: Comparison of the measured angle-dependent nonlocal voltage signals arising from thermally generated magnons in a freshly prepared sample and the same sample after one week of exposure to ambient air.

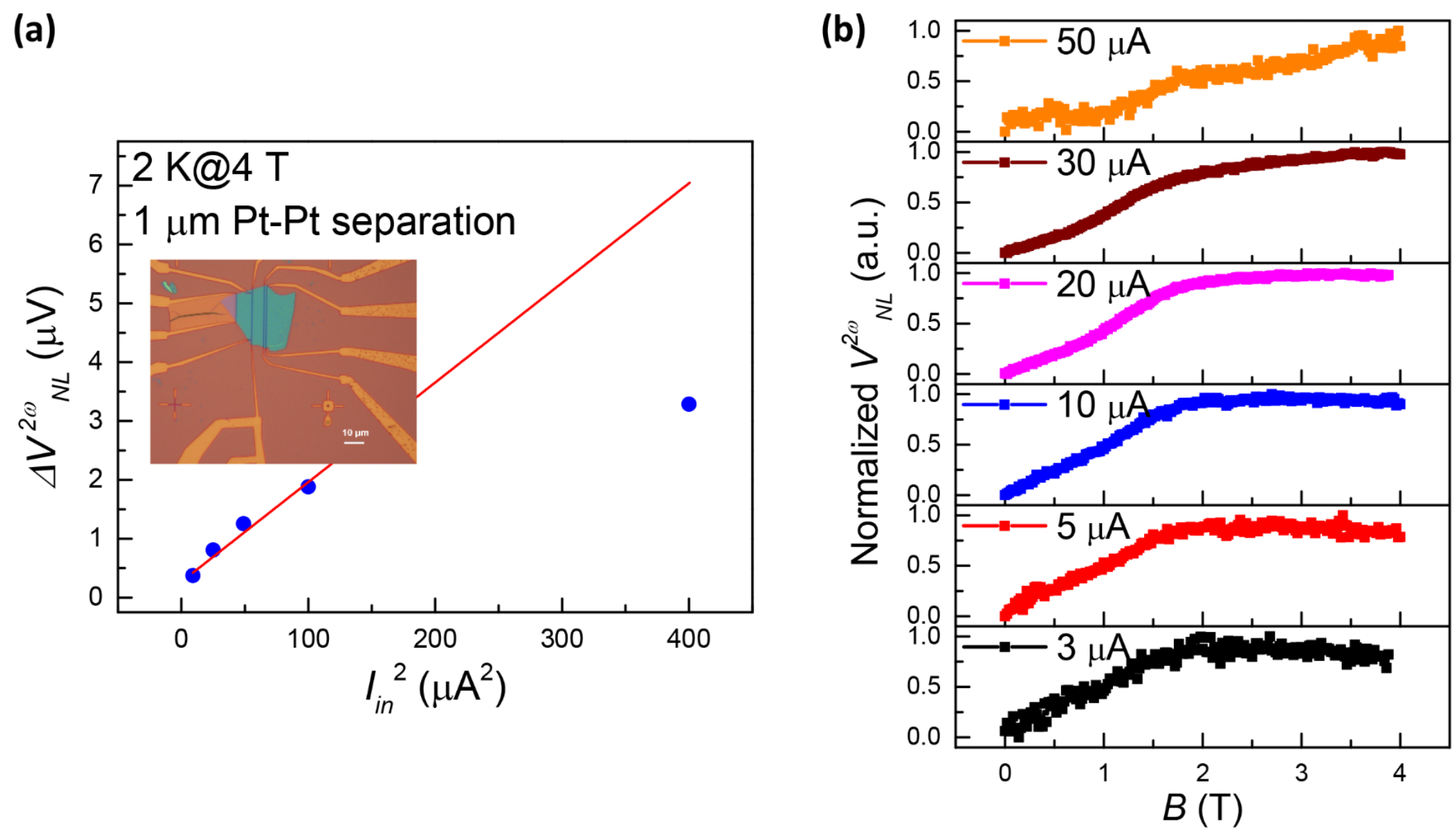

**Figure S7. The impact of high injection currents on the detected nonlocal voltages.** (a) Nonlocal voltage signal as function of injection currents, extracted from ADMR measurements of nonlocal magnon signals in Device 1. The inset displays an optical image of Device 1. The red line stands for a linear fit to the data points below 10 μA. (b) Normalized FDMR measurement of $V_{NL}^{2\omega}$ at different injection currents.

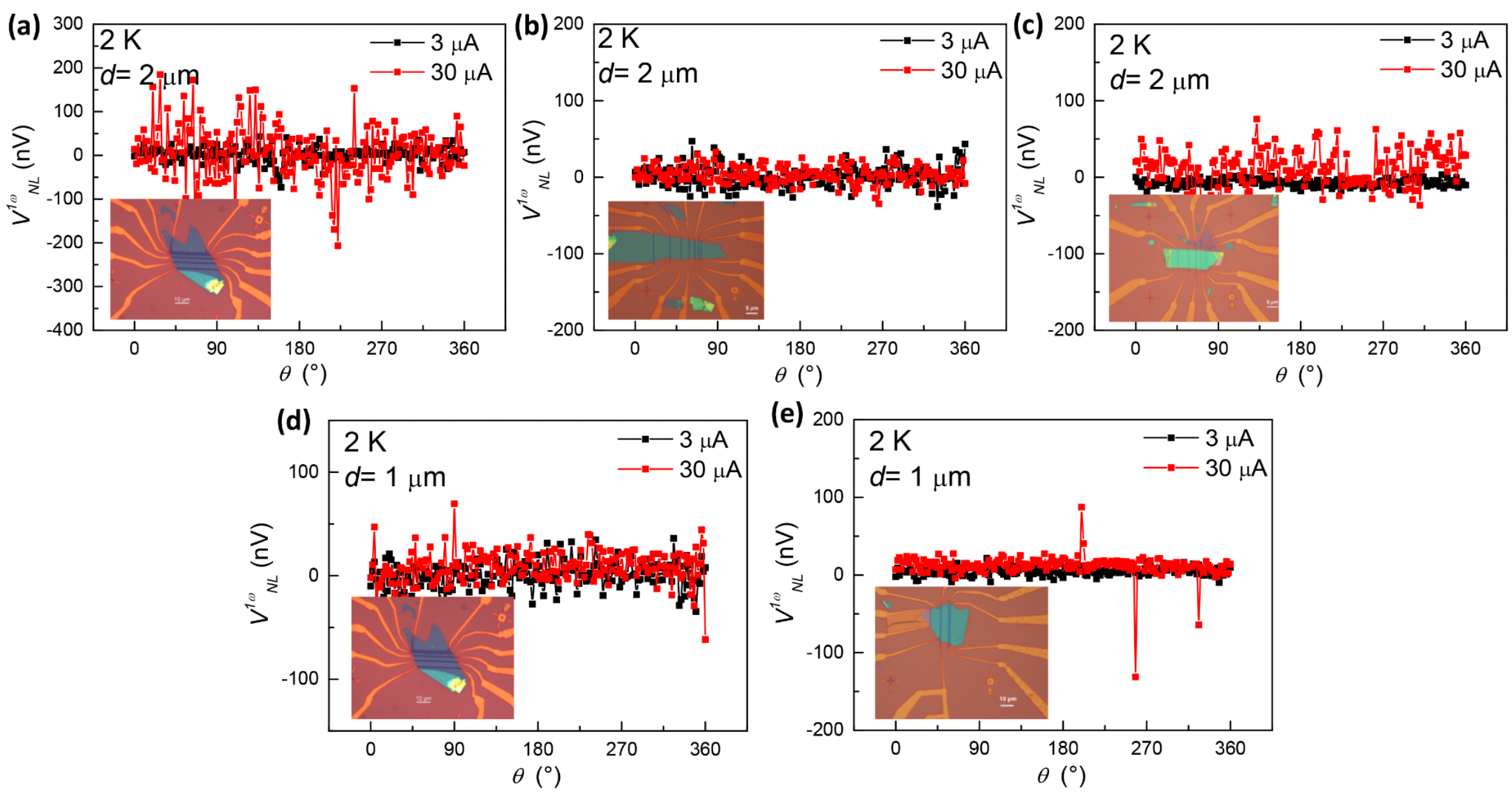

**Figure S8. Nonlocal ADMR measurement for the first-harmonic nonlocal voltages ($V_{NL}^{1\omega}$)** of $CrCl_3$ devices with thin thicknesses. The insets show the optical images of the corresponding devices. The collected data are from the devices with an injector-detector distance of 2 μm (a, c), and 1μm (d, e).

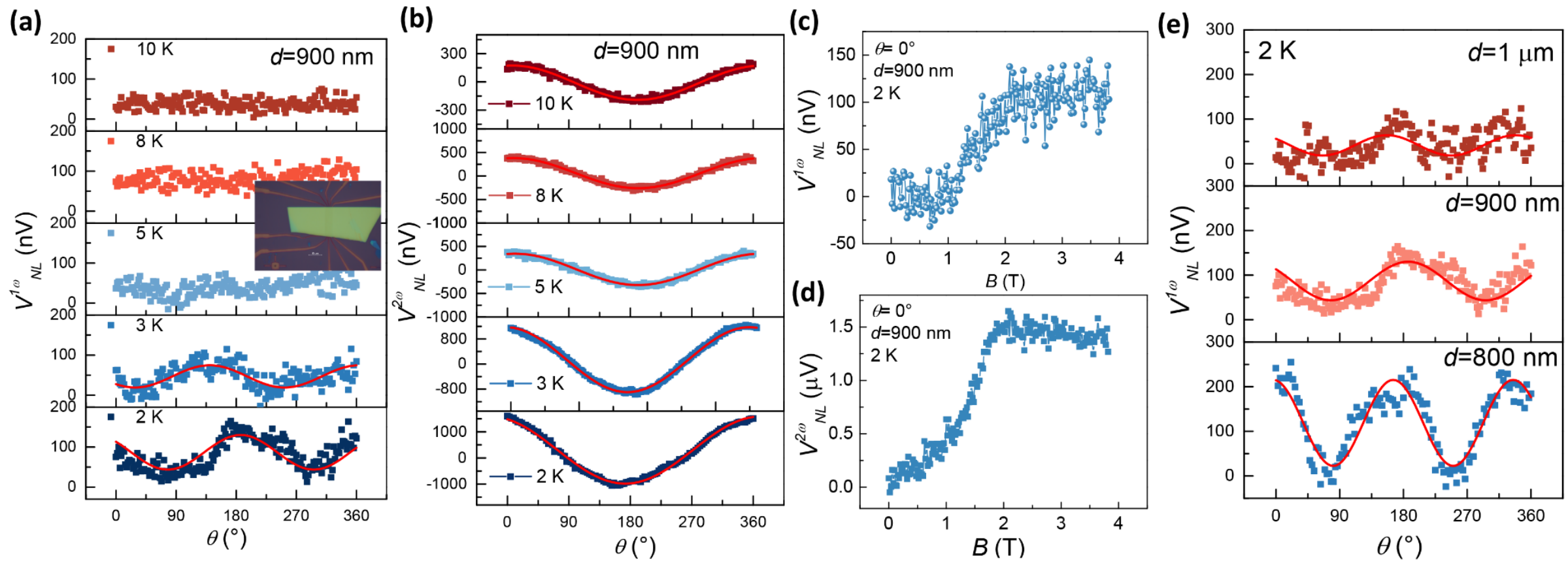

**Figure S9**. **Nonlocal voltage signal measurement of $CrCl_3$ devices with thickness larger than 100 nm.** (a, b) ADMR of $V_{NL}^{1\omega}$ and $V_{NL}^{2\omega}$ measured with *d*=900 nm under different temperatures. Red solid lines indicate fits to $cos^2\theta$ and $cos\theta$ dependence for $V_{NL}^{1\omega}$ and $V_{NL}^{2\omega}$, respectively. The inset shows the optical microscope image of the device. (c, d) FDMR of $V_{NL}^{1\omega}$ and $V_{NL}^{2\omega}$ at 2 K. (e) ADMR measurements of $V_{NL}^{1\omega}$ with different *d* at 2 K. The red solid lines are the fitted curves by the functions of $cos^2\theta$.

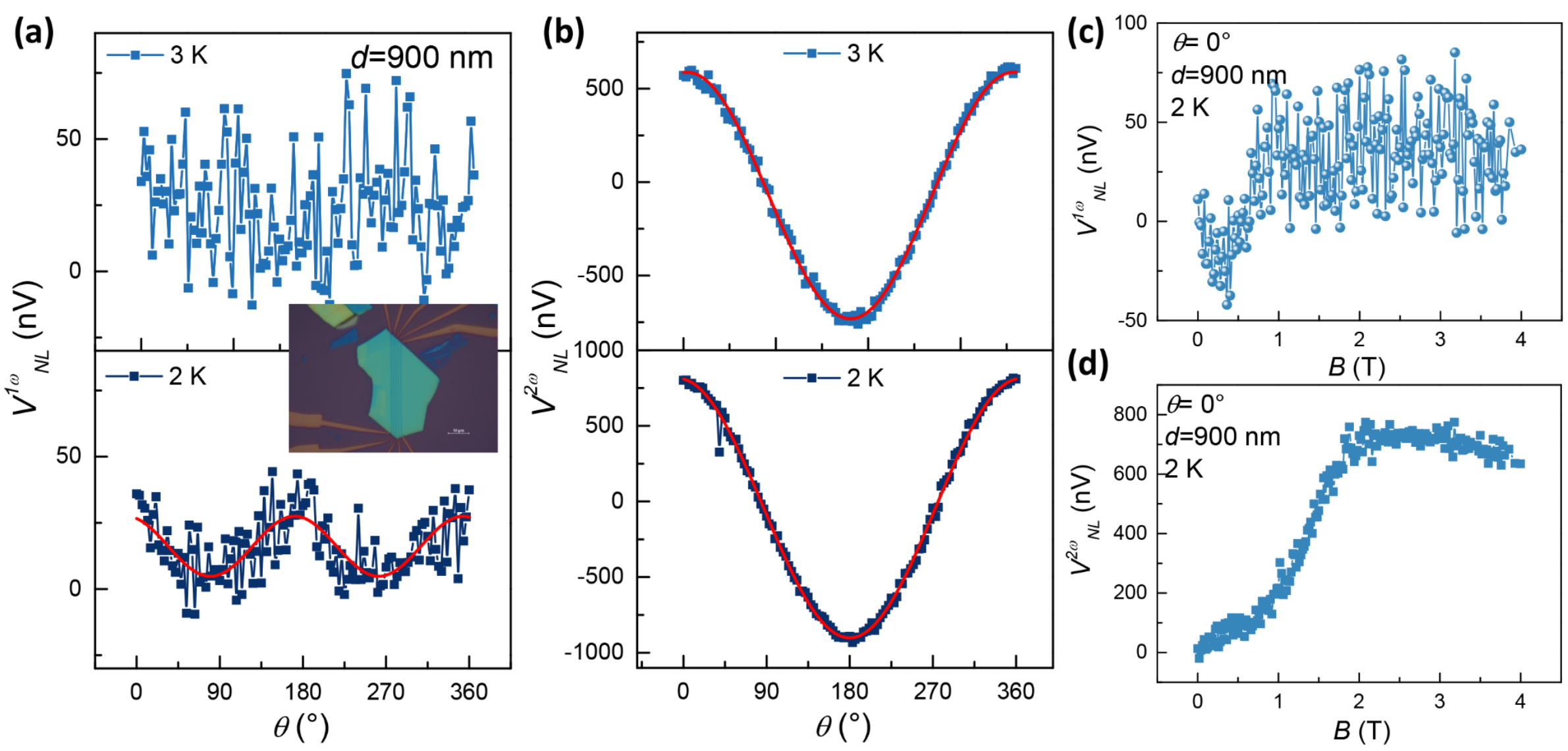

**Figure S10**. **Nonlocal voltage signal measurement of $CrCl_3$ devices with thickness around 80 nm.** (a, b) ADMR of $V_{NL}^{1\omega}$ and $V_{NL}^{2\omega}$ measured with *d*=900 nm under different temperatures. Red solid lines indicate fits to $cos^2\theta$ and $cos\theta$ dependence for $V_{NL}^{1\omega}$ and $V_{NL}^{2\omega}$, respectively. The inset shows the optical microscope image of the device. (c, d) FDMR of $V_{NL}^{1\omega}$ and $V_{NL}^{2\omega}$ at 2 K.

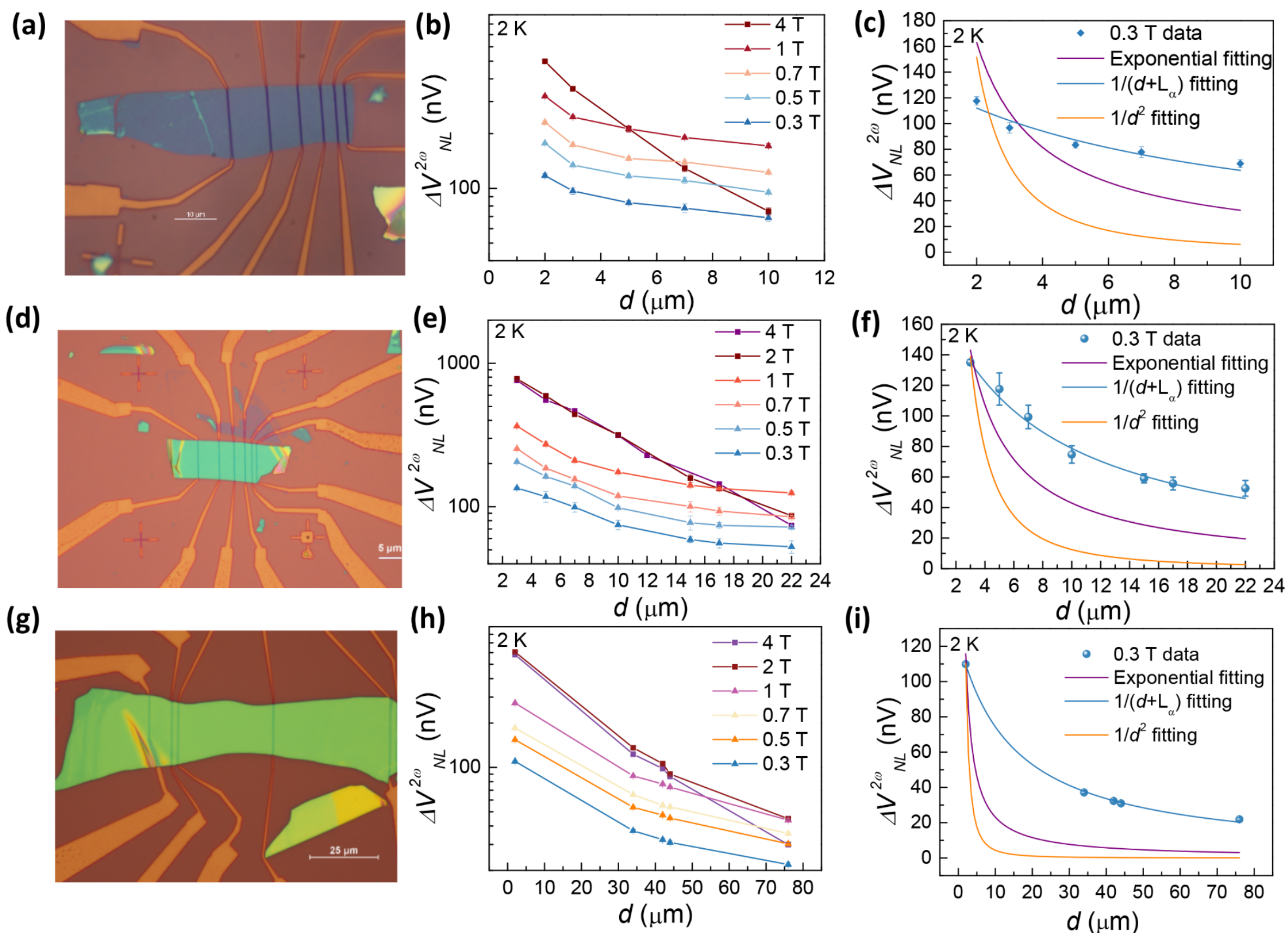

**Figure S11. Experimental evidence of superfluid spin transport in other $CrCl_3$ devices.** (a, d, g) Optical microscope images of the fabricated lateral device with Device 3 (a), Device 5 (d), and Device 6 (d). (b, e, h) Logarithmic plot of $\Delta V_{NL}^{2\omega}$ as a function of $d$ at various applied magnetic field strengths in Device 3 (b), Device 5 (e), and Device 6 (h). (c, f, i) Comparison of the fitting of $\Delta V_{NL}^{2\omega}$ versus $d$ at 0.3 T using different models in Device 3 (c), Device 5 (f), and Device 6 (i).

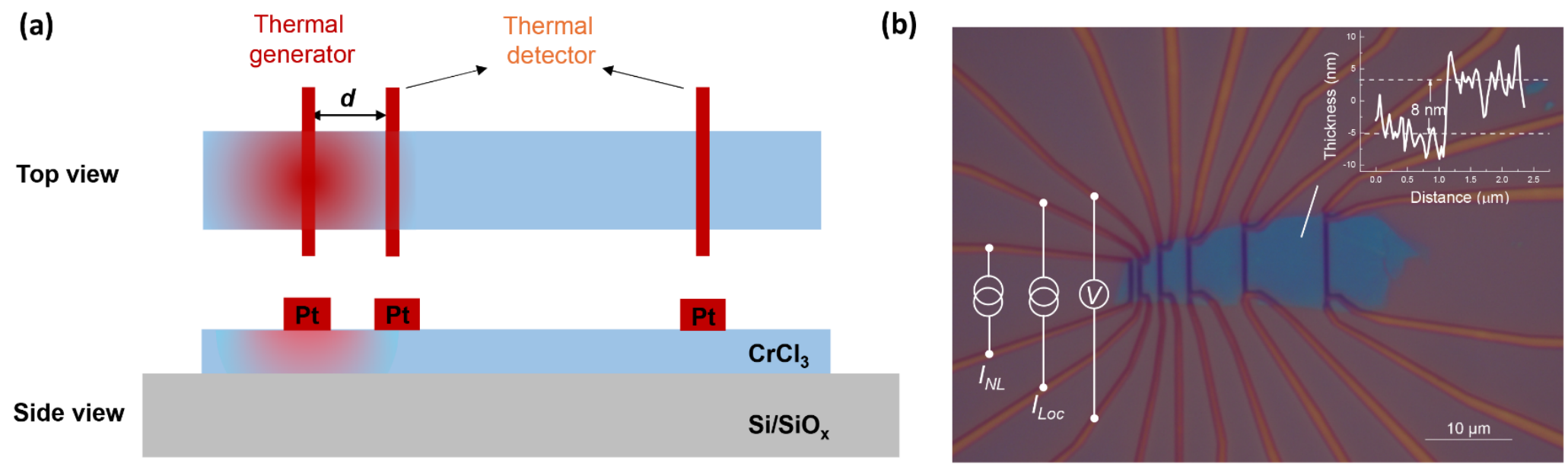

**Figure S12**. **Temperature profile measurement setup in the $CrCl_3$ nonlocal device.** (a) Schematic illustration of the experimental setup for measuring the temperature profile along the $CrCl_3$ channel. (b) Optical image of the fabricated nonlocal $CrCl_3$ device. Multiple Pt electrodes, each configured for four-probe resistance measurements, are positioned at different locations along the $CrCl_3$ channel, where the electrode separations ($d$) range from 1 to 9 μm. The inset line profile shows the height variations measured along the white solid line from the atomic force microscopy measurement. The $CrCl_3$ flake has a thickness of 8 nm, comparable to that of the device investigated in the main text.

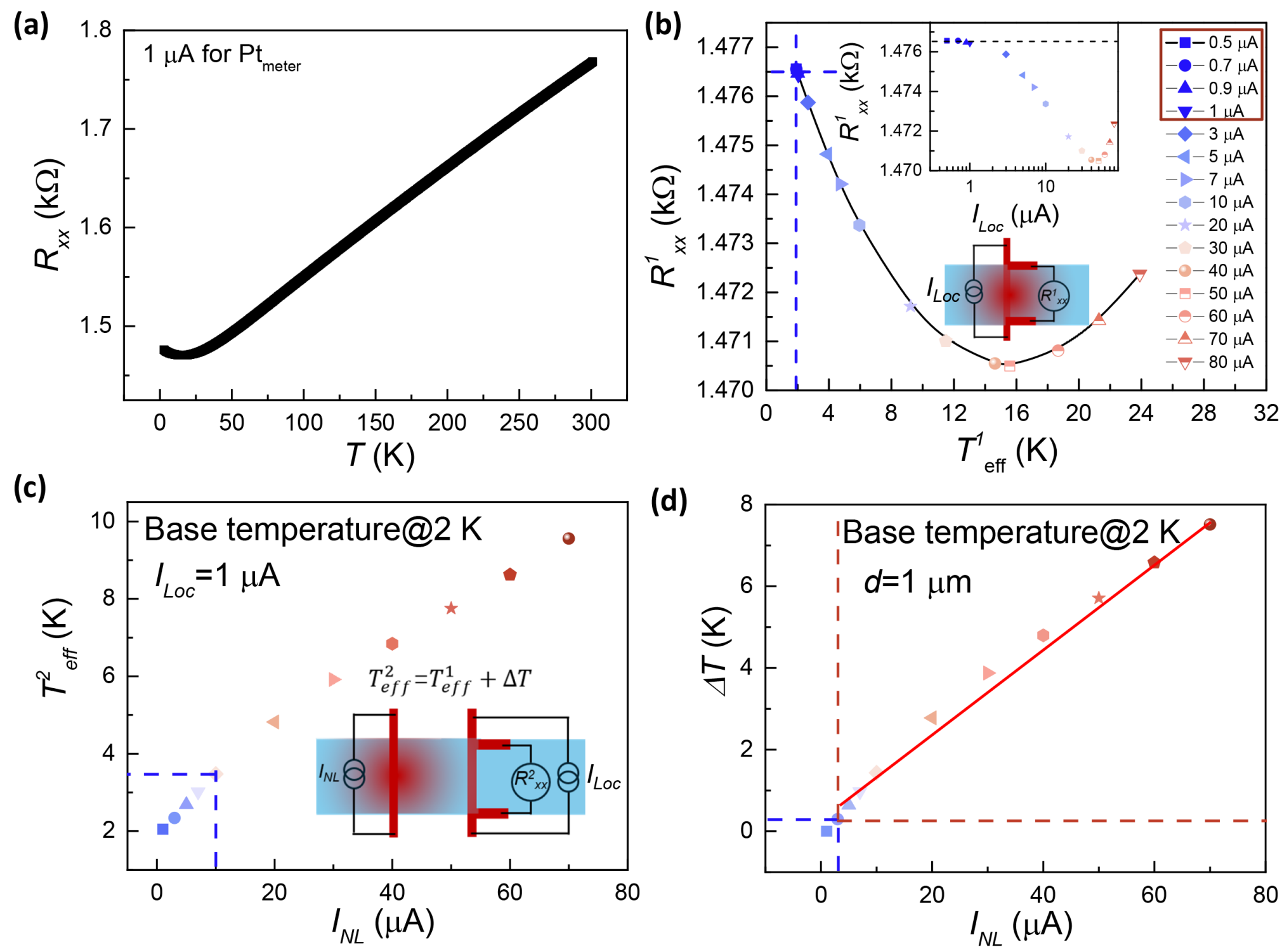

**Figure S13. Temperature profile measurements based on four-point resistance thermometry at $d$= 1 μm.** (a) Temperature-dependent resistance of a Pt electrode, used as the thermometer ($Pt_{meter}$), measured using a four-probe configuration. (b) Calibration curve relating the resistance ($R^1_{xx}$) of $Pt_{meter}$ to its effective temperature ($T^1_{eff}$) under different detector currents ($I_{Loc}$) at a base temperature of 2 K. The red square highlights the low-current regime in which the Pt resistance remains essentially unchanged, indicating negligible self-heating from the $I_{Loc}$. (c) Effective temperature $T^2_{eff}$ at the $Pt_{meter}$ position measured while varying the nonlocal injection current ($I_{NL}$) applied to the neighboring Pt injector ($Pt_{inj}$). (d) Temperature change ($\Delta T$) at the $Pt_{meter}$ position as a function of $I_{NL}$ applied to $Pt_{inj}$ for an electrode separation of $d$=1 μm. The blue and red dashed regions indicate negligible and detectable temperature changes within the equipment measurement error range, respectively. The red solid line serves as a guide to the eye.

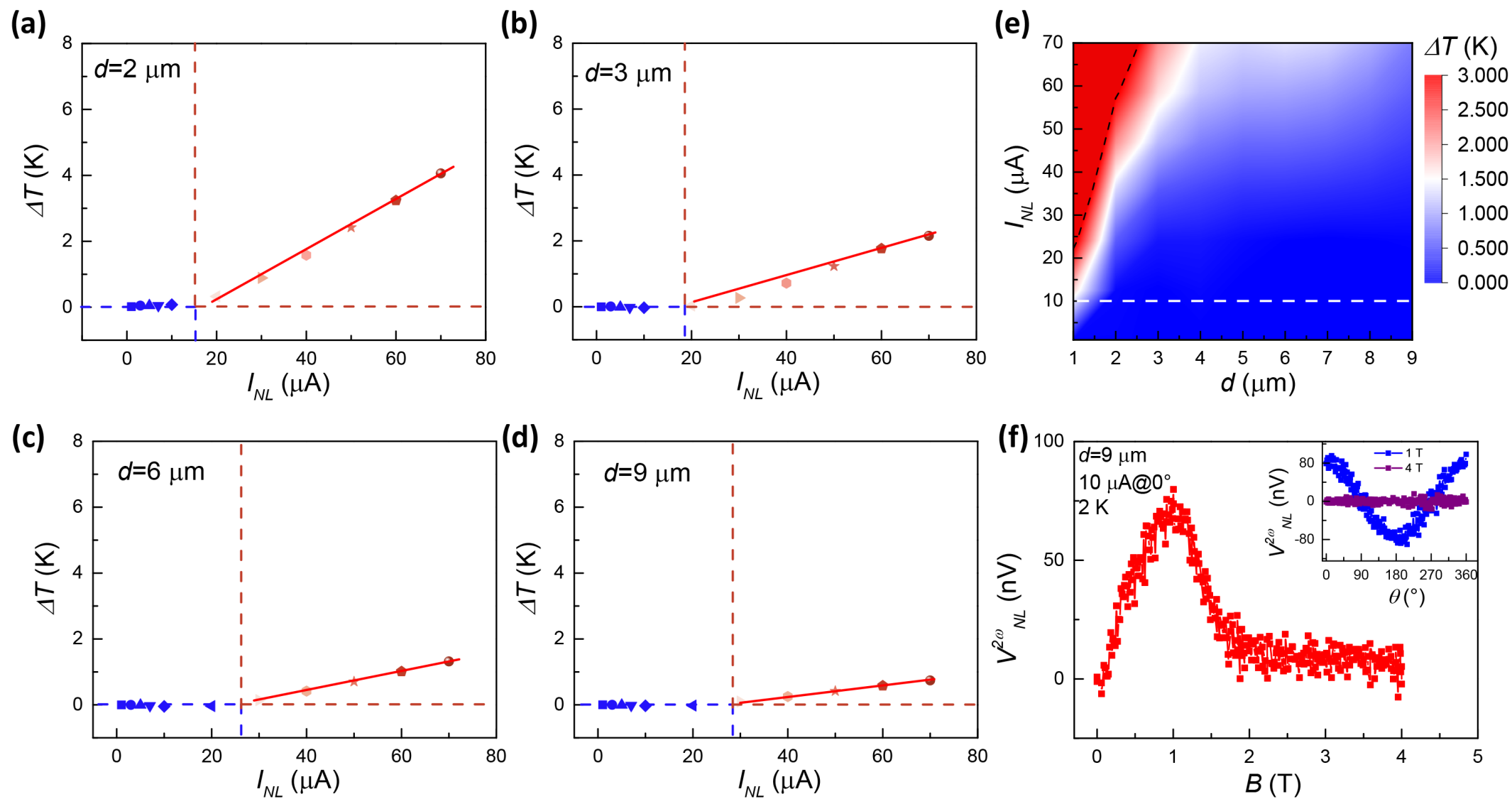

**Figure S14. Temperature profile and a nonlocal magnon signal measurements for longer *d* in an 8-nm-thick $CrCl_3$ device at a base temperature of 2 K.** $\Delta T$ at the $Pt_{meter}$ position as a function of the $I_{NL}$ applied to $Pt_{inj}$, with electrode separations of $d$=2 μm (a), 3 μm (b), 6 μm (c), and 9 μm (d). (e) Temperature-change map at $Pt_{meter}$ as a function of $d$ and $I_{NL}$. The white dashed line marks the $I_{NL}$=10 μA baseline used in the main-text measurement. The region above the black dashed line corresponds to $\Delta T$>3 K. (f) Magnetic-field- and angle-dependent second-harmonic nonlocal spin signal measured at $d$=9 μm.

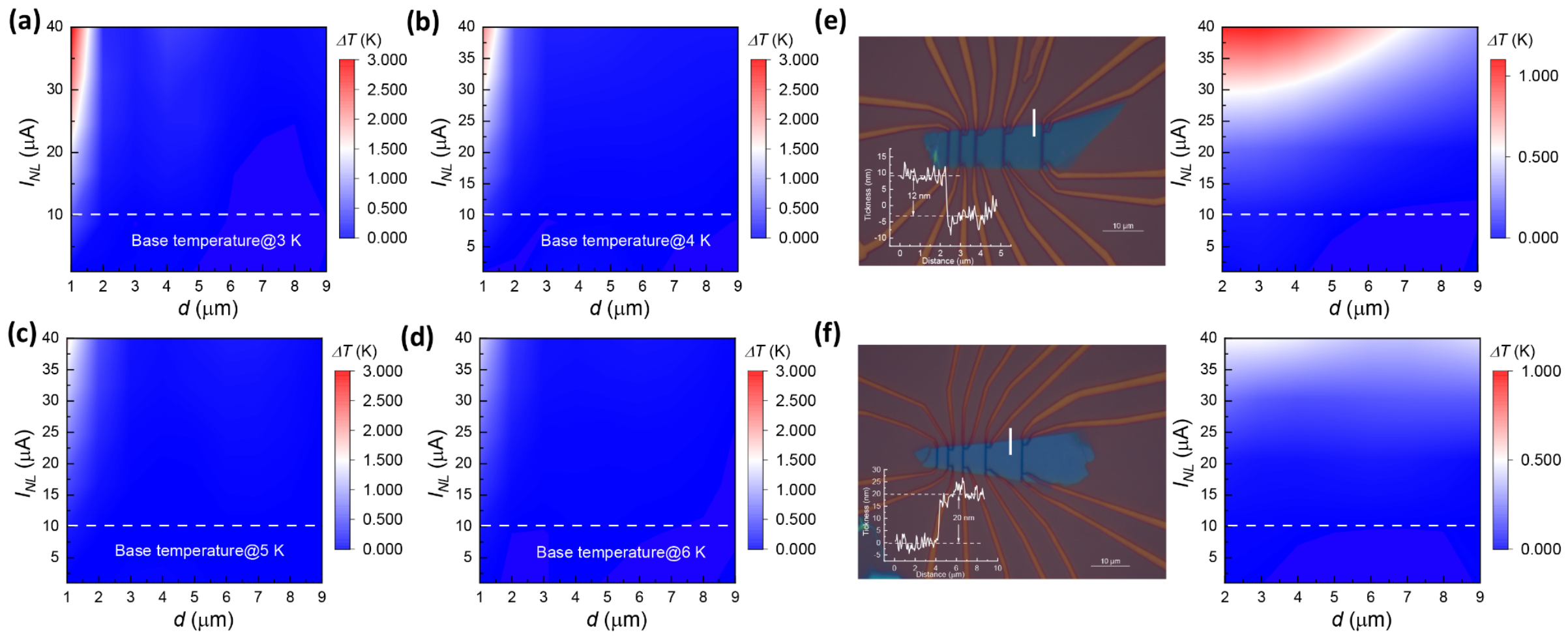

**Figure S15. Temperature profile measurement at different base temperatures and other $CrCl_3$ samples.** (a)-(d) Temperature-change map at $Pt_{meter}$ as a function of $d$ and $I_{NL}$ in the 8-nm-thick $CrCl_3$ device at other base temperatures, 3 K (a), 4 K (b), 5 K (c), and 6 K (d). (e) and (f) Temperature-change map at $Pt_{meter}$ as a function of $d$ and $I_{NL}$ in the 12-nm- (e) and 20-nm-thick (f) $CrCl_3$ devices measured at the base temperature of 2 K. The left panels show the optical images of the devices. The inset line profiles present the height variations measured along the white solid lines from the atomic force microscopy measurement. All the white dashed lines in temperature-change maps mark the $I_{NL}$=10 μA baseline used in the main-text measurement.

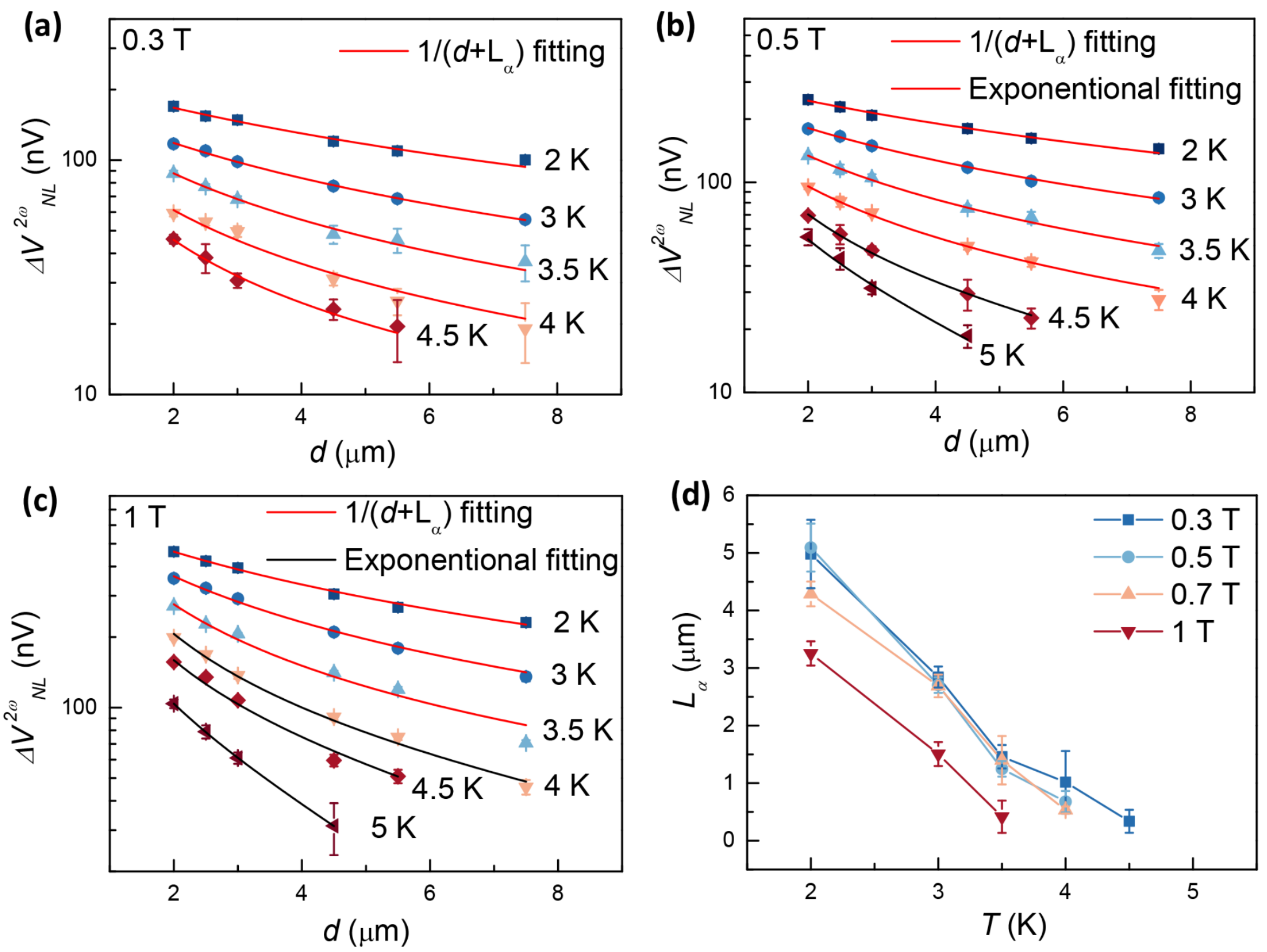

**Figure S16. Superfluid spin transport in Device 2 at different temperatures and fields.** (a)-(c) Semilogarithm ic plot of $\Delta V_{NL}^{2\omega}$ as a function of $d$ at various temperatures for fixed $B$= 0.3 T (a), 0.5 T (b) and 1 T (c) in Device 2. The red and black solid lines represent the fitted curves by the superfluid spin transport model of Eq. 1 and the conventional exponential decay function of Eq. 2 in the main text. (d) Fitted $L_\alpha$ values as function of temperature at different $B$.

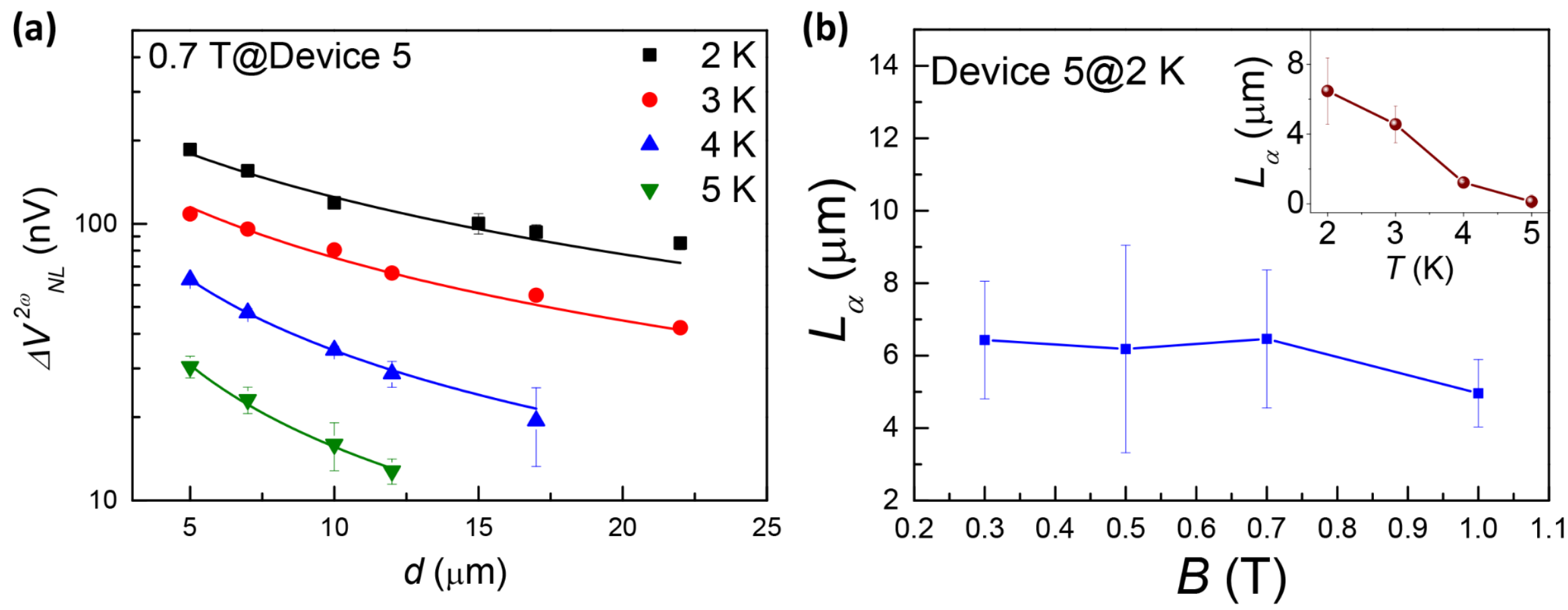

**Figure S17. Superfluid spin transport in Device 5 at different temperatures and fields.** (a) $\Delta V_{NL}^{2\omega}$ versus *d* at 0.7 T and different temperatures for Device 5. The solid lines represent the fitted curves by the superfluid spin transport equation (1) described in the main text. (b) Fitted $L_\alpha$ values as function of field strength at 2 K for Device 5. The inset shows the temperature dependent $L_\alpha$ values with a fixed field strength of 0.7 T.

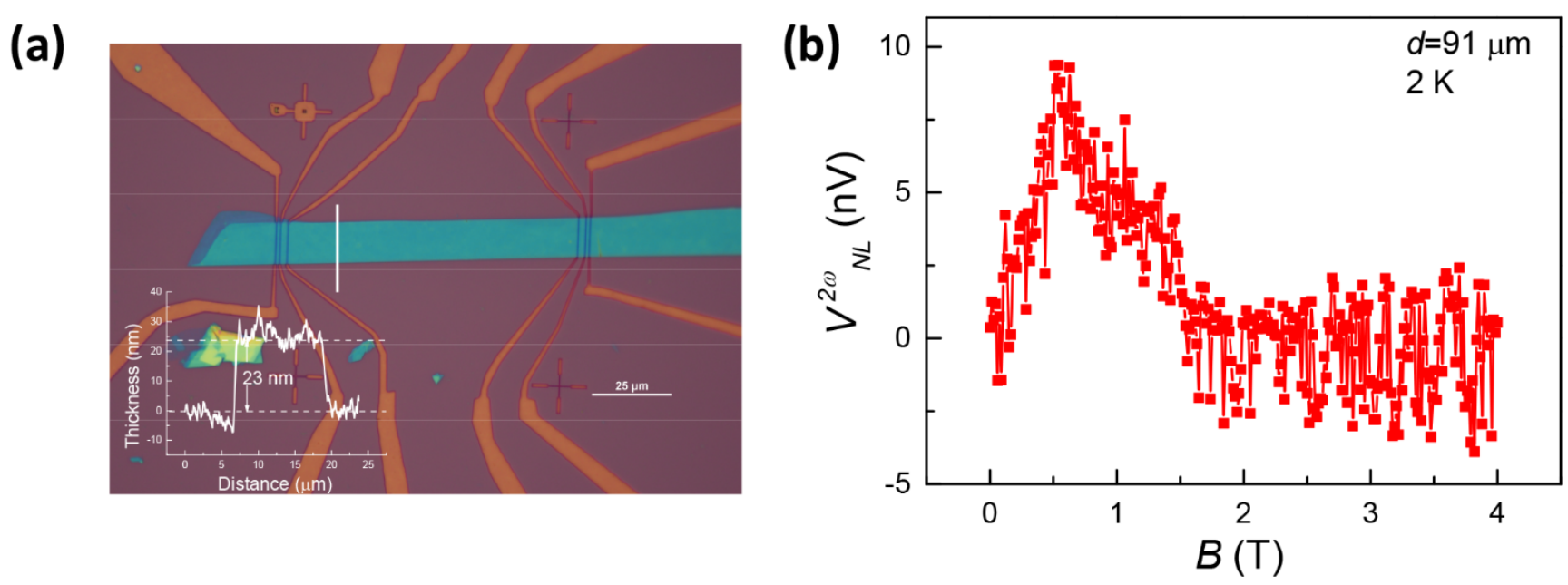

**Figure S18. Superfluid spin transport in $CrCl_3$ with ultralong distance.** (a) Optical image of a 23-nm-thick nonlocal $CrCl_3$ device. The inset line profile shows the height variations measured along the white solid line from the atomic force microscopy measurement. (b) FDMR of $V_{NL}^{2\omega}$ of this device measured at *d*=91 μm and 2 K.

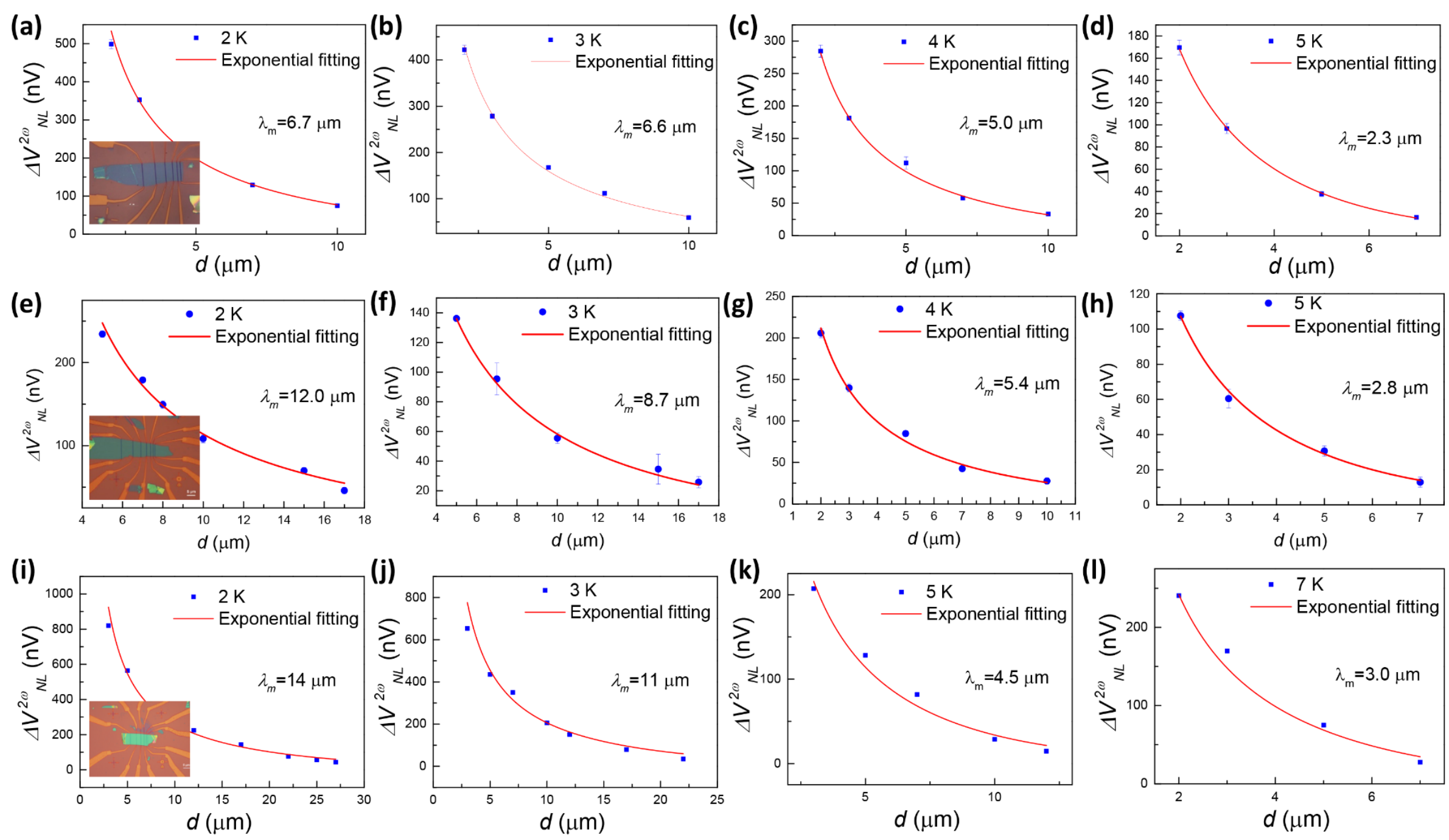

**Figure S19. The decaying dynamics of conventional thermally-driven magnon currents in Device 3** at 2 K (a), 3 K (b), 4 K (c), 5 K (d), Device 4 at 2 K (e), 3 K (f), 4 K (g), 5 K (h), and Device 5 at 2 K (i), 3 K (j), 5 K (k), 7 K (l). The solid red lines represent fitting curves by an exponential decay Eq. (2) described in the main text. $\lambda_m$ is the extracted magnon diffusion length from the exponential fitting. The insets of figures (a, e, i) show the optical images of Device 3, Device 4, and Device 5, respectively. All the data points are obtained from the ADMR measurement of $V_{NL}^{2\omega}$ signals at 4 T and 10 μA.

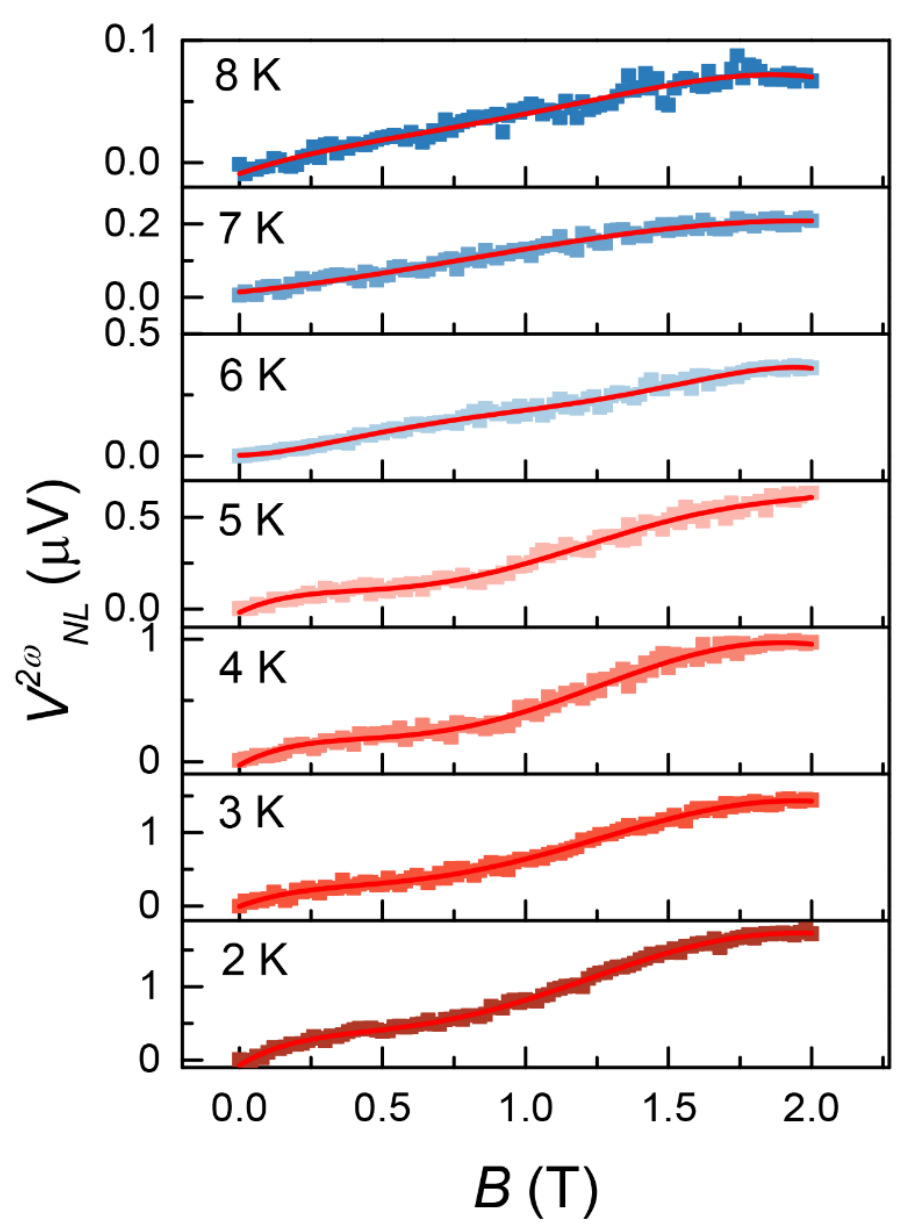

**Figure S20**. Temperature-dependent FDMR of $V_{NL}^{2\omega}$ for Device 1 at $d$=1 μm. The solid red lines serve as guide to the eye.

**Table S1. Comparison of our data with the reported nonlocal thermal magnon currentsand magnon diffusion length in other vdW or 3D MIs**

| | | $V_{NL}^{2\omega}/I^2$ (V/A$^2$) | $L_{eff}$ (μm) | $V_{NL}^{2\omega}/I^2$ (V/A$^2$ ·m) | $\lambda_m$ (μm) |
|---|---|---|---|---|---|
| vdW AFI | $CrCl_3$ | 6247 | 20.5 | 3.04 x10$^8$ | 14.1±2.9 |
| | $MnPS_3$ | 146 | 28.2 | 5.17 x 10$^6$ | 4.7±1.1 [17] |
| | $CrPS_4$ | 352 | 14.49 | 2.43 x 10$^7$ | 0.8 [18, 19] |
| | $FePS_3$ | 2.4 | 21.2 | 1.13 x 10$^5$ | 1.3±0.5 [20] |
| | $NiPS_3$ | 17 | 11.54 | 1.47 x 10$^6$ | 1.1±0.06 [21] |
| 3D AFI | α-$Fe_2O_3$ | 21.04 | 80 | 2.63 x 10$^5$ | 9±2 [24] |
| | $Cr_2O_3$ | 31.68 | 145 | 2.18 x 10$^5$ | 0.5 [25] |
| | NiO | -- | -- | -- | 0.0035-0.0045 [26] |
| 3D FI | YIG | 72.6 | 12.5 | 5.81x 10$^6$ | 8.7±0.8 [13] |
| | EuS | 639 | 70 | 9.13x 10$^6$ | 0.14 [27] |
| | $Fe_3O_4$ | -- | -- | -- | 0.034±0.008 [28] |

$L_{eff}$: the effective contact length of the Pt electrode on the MIs.

## Supplementary references

1 Sonin, E. B. Spin currents and spin superfluidity. *Adv. Phys.* **59**, 181-255 (2010).

2 Takei, S., Halperin, B. I., Yacoby, A. & Tserkovnyak, Y. Superfluid spin transport through antiferromagnetic insulators. *Phys. Rev. B* **90**, 094408 (2014).

3 Tserkovnyak, Y., Bender, S. A., Duine, R. A. & Flebus, B. Bose-Einstein condensation of magnons pumped by the bulk spin Seebeck effect. *Phys. Rev. B* **93**, 100402 (2016).

4 Takei, S. & Tserkovnyak, Y. Superfluid Spin Transport Through Easy-Plane Ferromagnetic Insulators. *Phys. Rev. Lett.* **112**, 227201 (2014).

5 Flebus, B. *et al.* Two-Fluid Theory for Spin Superfluidity in Magnetic Insulators. *Phys. Rev. Lett.* **116**, 117201 (2016).

6 Gomez-Perez, J. M. *et al.* Strong Interfacial Exchange Field in a Heavy Metal/Ferromagnetic Insulator System Determined by Spin Hall Magnetoresistance. *Nano Lett.* **20**, 6815-6823 (2020).

7 Sagasta, E. *et al*. Tuning the Spin Hall Effect of Pt From the Moderately Dirty to the Superclean Regime. *Phys. Rev. B*. **94**, 060412 (2016).

8 Kapoor, L. N. *et al*. Observation of Standing Spin Waves in a Van Der Waals Magnetic Material. *Adv. Mater.* **33**, 2005105 (2021).

9 Sonin, E. B. Superfluid Spin Transport in Ferro-and Antiferromagnets. *Phys. Rev. B*. **99**, 104423 (2019).

10 MacNeill, D. *et al*. Gigahertz Frequency Antiferromagnetic Resonance and Strong Magnon-Magnon Coupling in the Layered Crystal $CrCl_3$. *Phys. Rev. Lett.* , **123**, 047204 (2019).

11 Xue, F., Hou, Y., Wang, Z., Wu, R. Two-Dimensional Ferromagnetic Van Der Waals CrC $l_3$ Monolayer with Enhanced Anisotropy and Curie Temperature. *Phys. Rev. B*., **100**, 224429 (2019).

12 Le, K. B. *et al.* Magnon-Phonon Interactions From First Principles. arXiv preprint, arXiv:2502.05385. (2025).

13 Cornelissen, L. J., Liu, J., Duine, R. A., Youssef, J. B. & van Wees, B. J. Long-Distance Transport of Magnon Spin Information in a Magnetic Insulator at Room Temperature. *Nat. Phys.* **11**, 1022-1026 (2015).

14 Catalano, S. *et al.* Spin Hall Magnetoresistance Effect from a Disordered Interface. *ACS Applied Materials & Interfaces* **14**, 8598-8604 (2022).

15 Cai, X. *et al.* Atomically Thin $CrCl_3$: An In-Plane Layered Antiferromagnetic Insulator. *Nano Lett.* **19**, 3993-3998 (2019).

16 Wang, Z. *et al.* Determining the phase diagram of atomically thin layered antiferromagnet $CrCl_3$. *Nat. Nanotechnol.* **14**, 1116-1122 (2019).

17 Xing, W. *et al.* Magnon Transport in Quasi-Two-Dimensional van der Waals Antiferromagnets. *Phys. Rev. X* **9**, 011026 (2019).

18 de Wal, D. K. *et al.* Long-distance magnon transport in the van der Waals antiferromagnet $CrPS_4$. *Phys. Rev. B* **107**, L180403 (2023).

19 Qi, S. *et al.* Giant electrically tunable magnon transport anisotropy in a van der Waals antiferromagnetic insulator. *Nat. Commun.* **14**, 2526 (2023).

20 Feringa, F., Vink, J. M., & Van Wees, B. J. Spin Nernst magnetoresistance for magnetization study of $FePS_3$. *Phys. Rev. B* **107**, 094428 (2023).

21 Yuan, P. *et al.* Unconventional Magnon Transport in Antiferromagnet $NiPS_3$ Induced by an Anisotropic Spin-Flop Transition. *Nano Lett.* **25**, 5350-5357 (2025).

22 De Wal, D. K., Zohaib, M., Van, Wees, B. J. Magnon Spin Transport in the Van Der Waals Antiferromagnet $CrPS_4$ for Non-Collinear and Collinear Magnetization. *Phys. Rev. B*. **110**, 174440 (2024).

23 Wimmer, T. *et al*. Observation of antiferromagnetic magnon pseudospin dynamics and the Hanle effect. *Phys. Rev. Lett.* **125**, 247204 (2020).

24 Lebrun, R. *et al.* Tunable long-distance spin transport in a crystalline antiferromagnetic iron oxide. *Nature* **561**, 222-225 (2018).

25 Muduli, P. *et al.* Local and nonlocal spin Seebeck effect in lateral Pt–$Cr_2O_3$–Pt devices at low temperatures. *APL Mater.* **9**, 021122 (2021).

26 Guo, C. Y. *et al.* Magnon valves based on YIG/NiO/YIG all-insulating magnon junctions. *Phys. Rev. B* **98**, 134426 (2018).

27 Aguilar-Pujol, M. X. *et al.* Magnon currents excited by the spin Seebeck effect in ferromagnetic EuS thin films. *Phys. Rev. B* **108**, 224420 (2023).

28 Venkat, G. *et al.* Magnon diffusion lengths in bulk and thin film $Fe_3O_4$ for spin Seebeck applications. *Phys. Rev. Mater.* **4**, 075402 (2020).